\documentclass[journal,draftclsnofoot,onecolumn,12pt]{IEEEtran}
\ifCLASSINFOpdf
\else
\fi
\usepackage{graphicx}
\usepackage{bm}
\usepackage{amsfonts}
\usepackage{amsmath}
\usepackage{amssymb}
\usepackage{times}
\usepackage{cite}
\usepackage{subfigure}
\usepackage{latexsym,bm,amsmath,amssymb} 
\usepackage{CJK}
\usepackage{hhline}

\usepackage{multirow} 
\usepackage{amsmath}
\usepackage{xcolor}
\usepackage[linesnumbered,ruled,vlined]{algorithm2e}
\usepackage{epstopdf}

\usepackage{geometry}
\begin{document}

%
\title{A Novel ISAC Transmission Framework based on Spatially-Spread Orthogonal Time Frequency Space Modulation}

\author{Shuangyang~Li,~\IEEEmembership{Student Member,~IEEE,} Weijie~Yuan,~\IEEEmembership{Member,~IEEE,} Chang~Liu,~\IEEEmembership{Member,~IEEE,}
Zhiqiang~Wei,~\IEEEmembership{Member,~IEEE,}
Jinhong~Yuan,~\IEEEmembership{Fellow,~IEEE,}
Baoming~Bai,~\IEEEmembership{Senior Member,~IEEE,}
and
Derrick Wing Kwan Ng,~\IEEEmembership{Fellow,~IEEE}
\vspace{-8mm}
}


\maketitle


\begin{abstract}
In this paper, we propose a novel integrated sensing and communication (ISAC) transmission framework based on the spatially-spread orthogonal time frequency space (SS-OTFS) modulation by considering the fact that communication channel strengths cannot be directly obtained from radar sensing.
We first propose the concept of SS-OTFS modulation, where the key novelty is the angular domain discretization enabled by the spatial-spreading/de-spreading.
This discretization gives rise to simple and insightful effective models for both radar sensing and communication, which result in simplified designs for the related estimation and detection problems. In particular, we design simple beam tracking, angle estimation, and power allocation schemes for radar sensing, by utilizing the special structure of the effective radar sensing matrix. Meanwhile, we provide a detailed analysis on the pair-wise error probability (PEP) for communication, which unveils the key conditions for both precoding and power allocation designs. Based on those conditions, we design a symbol-wise precoding scheme for communication based only on the delay, Doppler, and angle estimates from radar sensing, without the \emph{a priori} knowledge of the communication channel fading coefficients, and also introduce the power allocation for communication.
Furthermore, we notice that radar sensing and communication requires different power allocations. Therefore,
we discuss the performances of both the radar sensing and communication with different power allocations and show that the power allocation should be designed leaning towards radar sensing in practical scenarios.
The effectiveness of the proposed ISAC transmission framework is verified by our numerical results, which also agree with our analysis and discussions.
\end{abstract}


\IEEEpeerreviewmaketitle
\section{Introduction}
Confronting the severe spectrum congestion, future wireless communication networks are expected to operate on higher frequency bands~\cite{liu2020joint}, such as the millimeter wave (mmWave) band. However, a large portion of spectral resources in those bands has been preliminarily assigned to radar systems. For example, automotive radars and high-resolution imaging radars are usually operating on the mmWave band of $76$-$81$ GHz and $200$ GHz, respectively~\cite{choi2016millimeter}.
Therefore, it is crucial for future wireless networks to achieve a harmonious coexistence with radar systems in order to achieve the ultra-high throughput requirements~\cite{liu2020joint,liu2018toward,yuan2021integrated,liu2018mu}.
There are mainly two approaches to achieve the coexistence between wireless communication networks and radar systems. One is to design efficient interference management algorithms to maintain the functionalities of both radar sensing and communication, by suppressing the interference generated from each other~\cite{liu2020joint,liu2018mimo}. Another one is to consider an integrated sensing and communication (ISAC) system, where a well-designed waveform is transmitted for the purposes of both radar sensing and communication~\cite{liu2018toward,yuan2021integrated}. Compared with the interference management design, ISAC transmissions allow effective cooperation between the two functionalities and have also shown substantial potentials for various emerging applications~\cite{wymeersch20175g,yuan2021integrated,liu2018toward}. More interestingly, it has also been reported that a joint design between radar sensing and communication may potentially lead to performance improvements for both the functionalities~\cite{kobayashi2018joint}. Therefore, ISAC transmissions have been attracting substantial attention lately.

A key motivation for ISAC transmission designs is that both radar sensing and communication naturally have similar channel characteristics which can be exploited.
For example, let us consider a common downlink scenario in a mobile network, where the antennas for radar sensing and communication are co-located at the base station (BS).
It is not hard to notice that the physical channel between the BS and user equipments (UEs) is the same for both radar sensing and communication, despite the fact that the radar sensing is operated based on the received echoes at the BS after the \emph{round-trip} signal propagation, while the signal detection for communication is based on the \emph{one-way} transmission from the BS to UEs.
Note that all the transmitted signals are known to the radar, which can be used for the related sensing purposes. Hence,
radar sensing usually enjoys a much higher matched-filtering gain for parameter estimation compared to the channel estimation algorithms achieved in communication systems~\cite{yuan2021integrated}.
Therefore, it is wise to exploit the channel state information (CSI) obtained from the radar sensing to facilitate an effective communication design.
Furthermore, it is also worthwhile to notice that radar sensing carries out parameter estimations based on
the delay, Doppler, and angular features associated to resolvable paths, whose core idea aligns perfectly with the recently proposed orthogonal time frequency space (OTFS) modulation for communication transmission in future communication networks~\cite{Hadani2017orthogonal,Zhiqiang_magzine,wei2020transmitter}. Specifically, OTFS modulation relies on the exploration of the delay-Doppler (DD) domain symbol multiplexing and DD domain channel characteristics, which is different from the conventional time-frequency (TF) domain symbol multiplexing as adopted in the orthogonal frequency-division multiplexing (OFDM) modulation.
The DD domain symbol multiplexing enables the direct interactions between the information symbols and the DD domain channel, whose channel response can be potentially inferred from the radar estimates in practice~\cite{yuan2021integrated}.
The synergistic ecosystem established by needs of communication and radar sensing has motivated us to consider the ISAC transmission design based on OTFS modulation.
\begin{figure}
\centering
\includegraphics[width=0.5\textwidth]{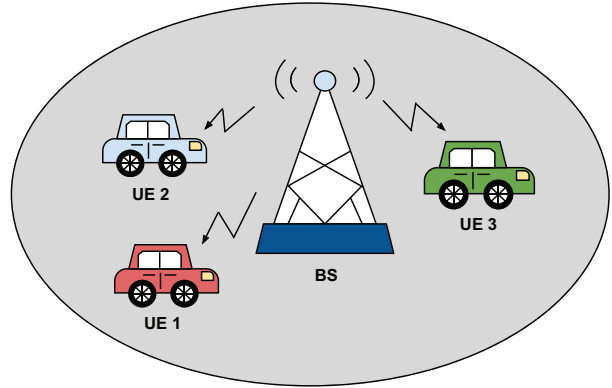}
\caption{A brief diagram of the considered ISAC transmission.}
\label{ISAC}
\vspace{-6mm}
\centering
\end{figure}

To unlock the potential of OTFS modulation-based ISAC transmissions, various lines of research works have been conducted in the literature. For example, the effectiveness of OTFS modulation for ISAC transmission has been evaluated in~\cite{Gaudio2020on}, where the authors have shown that the estimation error lower bounds for radar sensing can be achieved by using OTFS signals while maintaining a satisfactory communication performance. Then, this work has then been extended to the case of multiple-input multiple-output (MIMO)~\cite{gaudio2020hybrid}, where a hybrid digital-analog beamforming is devised for both radar sensing and communication. In addition, the author in~\cite{gaudio2020hybrid} have also developed an efficient maximum-likelihood (ML) algorithm to facilitate target detection and parameter estimation.
In addition, a novel OTFS-based matched-filter algorithm for target range and velocity estimation for radar has been proposed in~\cite{raviteja2019orthogonal}. Specifically, the proposed matched-filter algorithm takes advantages of the structures of DD domain effective channel matrix and has shown better estimation performance compared to the OFDM counterpart.
Furthermore, an ISAC-assisted OTFS system has been proposed in~\cite{yuan2021integrated}, where both uplink and downlink communications are considered. In particular, the authors proposed a novel DD domain channel estimation algorithm and introduced a message-passing based detection algorithm for uplink transmission. On the other hand, the downlink communication transmission is designed based on the CSI obtained from radar sensing, such that it can bypass the need of channel estimation and equalization at the receiver side.
Although the applications of OTFS modulation in ISAC transmissions have shown promising performances, they often rely on sophisticated beamforming schemes~\cite{yuan2021integrated,gaudio2020hybrid} that are designed according to the CSI at transmitter (CSIT). However, it should be noted that for some practical scenarios, such as co-located radar and communication antennas, channel fading coefficients for communication cannot be directly obtained from radar sensing. In specific, the strengths of channel fading coefficients for communication usually depend on the path loss and channel scatters ~\cite{hlawatsch2011wireless}, while the echo strengths for radar sensing also depend on the effective area of the radar receiving antenna and the radar cross section (RCS)~\cite{shrestha2008method}.
Therefore, there is generally a mismatch of the reflection strengths between the radar sensing and communication.
In other words, the path with the strongest echo power for radar sensing may not be the strongest path for communication. Consequently, if the beams for communication steer towards the strongest path indicated by radar sensing, the communication performance may degrade dramatically.

Considering the potential mismatch between radar sensing and communication, we propose a novel ISAC transmission framework based on spatially-spread OTFS (SS-OTFS) modulation. To facilitate the ISAC design, we introduce the concept of SS-OTFS modulation for the first time in the literature to further exploit the delay-Doppler-angular (DDA) domain channel characteristics. Compared to conventional MIMO-OTFS modulation, SS-OTFS applies the so-called ``spatial spreading" and ``spatial de-spreading" modules at the transmitter and receiver, respectively.  The key novelty of applying those modules is the \textbf{discretization} of the angular domain, which results in simple and insightful input-output relationships for both radar sensing and communication.
The most interesting feature of those relationships is that each antenna (pair) corresponds to a specific angle according to the \emph{angular resolution}. As such, it is possible to fully separate the multi-path effect, which enables efficient system designs that based only on estimates of the delays, Dopplers, AoAs, and radar reflection coefficients from radar sensing, without the \emph{a priori} knowledge of the fading coefficients of the communication channels.
The main contributions of this paper can be summarized as follows.
\begin{itemize}
\item We derive both the communication and radar models for SS-OTFS-enabled ISAC transmission. In particular, we show that the interference from spatial multiplexing can be approximately eliminated by spatial spreading and de-spreading with a sufficiently large number of antennas, which results in simple and insightful effective channel matrices for both radar sensing and communication.
\item Based on the radar sensing model, we develop simple beam tracking and AoA estimation algorithms by exploiting the special structure of the effective radar sensing matrix. We show that the transmitted beam width can be easily controlled by the power allocation among the antennas, which is independent from the precoding design. Furthermore, the AoA estimation can be straightforwardly implemented by checking the received power for different antennas, which is due to the discretization of the angular domain. Furthermore, we introduce the power allocation for radar sensing, which is designed to maximize the minimum power of the received echoes.

\item Based on the derived communication model, we analytically unveil the impacts of precoding matrices and power allocation on the pair-wise error probability (PEP). In particular, we show that the PEP is minimized when the \emph{equivalent codeword difference matrix} has a diagonal structure and the geometric mean of the allocated power associated to corresponding paths is maximized. Based on this finding, we develop our precoding design by introducing \emph{virtual} delay and Doppler indices to shape the equivalent codeword difference matrix, while we show that the equal power allocation can maximize the geometric mean.
\item We notice that radar sensing and communication require different power allocations. Therefore, we discuss the radar sensing and communication performances with respect to different power allocations. Based on our discussions and simulation results, we show that the power allocation should be designed leaning towards radar sensing in practical scenarios. Meanwhile, the effectiveness of the proposed ISAC framework has also been verified by our simulation results.
\end{itemize}

\emph{Notations:} The blackboard bold letters ${\mathbb{A}}$, ${\mathbb{C}}$, and ${\mathbb{E}}$ denote the energy-normalized constellation set, the complex number field, and the expectation operator, respectively; $\textrm{det}(\cdot)$, $\textrm{Tr}(\cdot)$, $\textrm{vec}(\cdot)$, and ${\left\| {\cdot} \right\|_{\rm{F}}}$ denote the determinant, the trace, the vectorization, and the Frobenius norm operations, respectively; $\textrm{span}(\cdot)$ denotes the span of a set; $\textrm{diag}{\{\cdot\}}$ denotes a diagonal matrix or a block diagonal matrix; ``$ \otimes $" denotes the Kronecker product operator; ${{{\bf{F}}_N}}$, ${{{\bf{I}}_N}}$, and ${{{\bf{0}}_N}}$ denote the discrete Fourier transform (DFT) matrix, the identity matrix, and an all-zero matrix
of size $N\times N$, respectively; ${\left[ {\cdot} \right]_N}$ denotes the modulo-$N$ operation; ${\left( {\cdot} \right)_{\min }}$ denotes the minimum value; $\Pr \left\{ {\cdot} \right\}$ denotes the probability of an event; ${f_{{\rm{PDF}}}}\left( x \right)$ denotes the power density function (PDF) of a random variable $x$.
\section{System Model}
Without loss of generality, let us consider an ISAC system in a mobile network, where one BS broadcasts a common message to $K$ randomly distributed UEs within the service coverage and senses the radar-related information of the UEs based on the received echoes. In particular, we consider a multiple-input single-output (MISO) case, where the BS is equipped $N_{\rm BS}$ antennas while each UE has only one antenna. We assume that the system operates in an open area as shown in Fig.~\ref{ISAC}, where there are $P$ independent resolvable paths between the BS and each UE{\footnote {We note that this scenario is commonly considered for practical systems, some examples and explanations can be found in~\cite{liu2020joint}.}}.
\subsection{Transmitter Structure}
\begin{figure}
\centering
\includegraphics[width=0.7\textwidth]{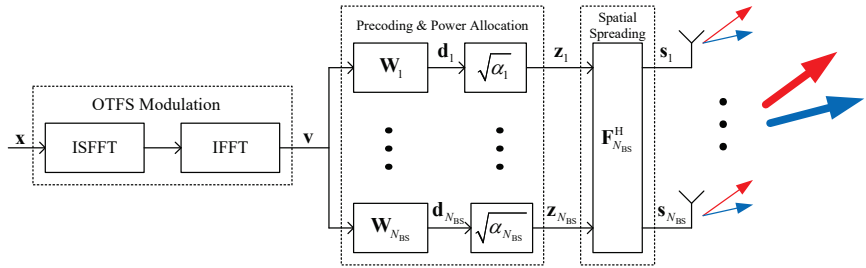}
\caption{The block diagram of the SS-OTFS-enabled ISAC transmitter, where the two arrows represent different beam directions.}
\label{transmitter_model}
\vspace{-3mm}
\centering
\end{figure}

\begin{figure}
\centering
\includegraphics[width=0.7\textwidth]{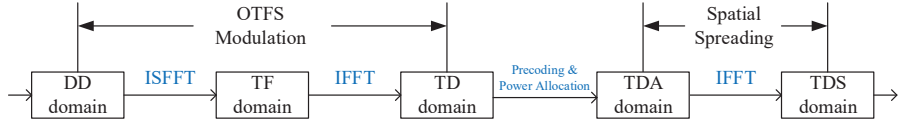}
\caption{The domain transformations the SS-OTFS enabled ISAC transmitter.}
\label{domain_transform}
\vspace{-5mm}
\centering
\end{figure}

Without loss of generality, let us consider the SS-OTFS-enabled ISAC transmitter structure as shown in Fig.~\ref{transmitter_model}.
Let ${\bf{X}} \in { {\mathbb A}^{M \times N }}$ be the DD domain transmitted symbol matrix (broadcast information message) of size $M \times N$, where $M$ denotes the number of orthogonal subcarriers and $N$ denotes the number of time slots, respectively.
Let ${\Delta f}$ and $T$ be the subcarrier spacing and the time slot duration, respectively.
By performing the inverse symplectic finite Fourier
transform (ISFFT) and IFFT to ${\bf{X}}$, the time-delay (TD) domain transmitted symbol matrix ${\bf{V}} \in { {\mathbb A}^{M \times N }}$ of size $M \times N$ can be obtained.
For simplicity, we consider the vector form representation of OTFS transmissions according to~\cite{Raviteja2019practical}. Let ${\bf{x}} \buildrel \Delta \over= {\rm{vec}}\left( {\bf{X}} \right)$ and  ${\bf{v}} \buildrel \Delta \over= {\rm{vec}}\left( {\bf{V}} \right)$ be the DD domain and the TD domain transmitted symbol vectors of length $MN$, respectively. Then, we have~\cite{Raviteja2019practical,li2021performance}
\begin{align}
{\bf{v}}= \left( {{\bf{F}}_N^{\rm{H}} \otimes {{\bf{I}}_M}} \right){\bf{x}}. \label{TD_transmitted_symbol_vec}
\end{align}
After obtaining the TD domain broadcast message ${\bf{v}}$, the BS multiplexes the message onto each antenna with respect to the precoding matrices{\footnote{It is worth noticing that the conventional precoding matrix for narrow band multiple-input systems is of size $N_{\rm BS}\times N_{\rm BS}$. However, the transmitted signal on each antenna is generally a wideband signal in the considered system. Therefore, we propose to apply precoding to each antenna's transmitted signal in order to combat the multi-path interference, while apply spatial spreading to combat the interference after spatial multiplexing.}} $\left\{ {{{\bf{W}}_1},{{\bf{W}}_2},...,{{\bf{W}}_{{N_{{\rm{BS}}}}}}} \right\}$ and we have
\begin{align}
{{\bf{d}}_{n_{\rm t}}} = {{\bf{W}}_{n_{\rm t}}}{{\bf{v}}}, \quad \forall {n_{\rm{t}}} \in \left\{ {1,2,...,{N_{{\rm{BS}}}}} \right\}, \label{signals_before_PA}
\end{align}
where the precoding matrix ${{\bf{W}}_{n_{\rm t}}}$ is of size $MN \times MN$ and it  has a normalized energy with respect to the length of the message length, i.e., ${\left\| {{{\bf{W}}_{{n_{\rm{t}}}}}} \right\|_{\rm{F}}} = MN$, for $1 \le {n_{\rm{t}}} \le {N_{{\rm{BS}}}}$. In particular, we restrict ourselves to only consider the symbol-wise precoding such that each row/column of ${{\bf{W}}_{n_{\rm t}}}$ only has one non-zero element and ${{\bf{W}}_{{n_{\rm{t}}}}}{\bf{W}}_{{n_{\rm{t}}}}^{\rm{H}} = {{\bf{I}}_{MN}}$, for $1 \le {n_{\rm{t}}} \le {N_{{\rm{BS}}}}$.
Based on~\eqref{signals_before_PA}, we allocate power to each antenna's signal, such that the transmitted symbol vector ${\bf z}_{n_{\rm t}}$ after power allocation{\footnote{For the ease of presentation, we henceforth use the term energy and power interchangeably, without raising ambiguities.}} for the $n_{\rm t}$-th antenna is given by
\begin{align}
{{\bf{z}}_{n_{\rm t}}} = \sqrt {{\alpha _{{n_{\rm{t}}}}}} {{\bf{d}}_{n_{\rm t}}}, \label{signals_before_BF}
\end{align}
where $\alpha_{n_{\rm t}}$, for $1 \le {n_{\rm{t}}} \le {N_{{\rm{BS}}}}$, is the allocated power for the ${n_{\rm{t}}}$-th antenna, and $\sum\nolimits_{{N_{\rm{t}}} = 1}^{{N_{{\rm{BS}}}}} {{\alpha _{{n_{\rm{t}}}}}}  = {\alpha _{{\rm{total}}}}$ with ${\alpha _{{\rm{total}}}}$ being the total transmit power budget.
Let us define the transmitted symbol vector before and after power allocation by ${\bf{d}} \buildrel \Delta \over = {\left[ {{\bf{d}}_1^{\rm{H}},{\bf{d}}_2^{\rm{H}},...,{\bf{d}}_{{N_{{\rm{BS}}}}}^{\rm{H}}} \right]^{\rm{H}}}$ and ${\bf{z}} \buildrel \Delta \over = {\left[ {{\bf{z}}_1^{\rm{H}},{\bf{z}}_2^{\rm{H}},...,{\bf{z}}_{{N_{{\rm{BS}}}}}^{\rm{H}}} \right]^{\rm{H}}}$, respectively.
Then, it can be shown that
\begin{align}
{\bf{z}} = \left( {{\bm \alpha}  \otimes {{\bf{I}}_{MN}}} \right){\bf{d}}, \label{power_allocation_vec_form}
\end{align}
where ${\bm \alpha}\buildrel \Delta \over = {\rm diag}\left\{ {\sqrt{\alpha _1},\sqrt{\alpha _2},...,\sqrt{\alpha _{{N_{{\rm{BS}}}}}}} \right\}$ is the diagonal power allocation matrix of size $N_{\rm BS}\times N_{\rm BS}$.
Let us rearrange the transmitted symbol vectors for different antennas into a matrix $\bf Z$ of size $MN \times N_{\rm BS}$ based on ${\bf{z}} = {\rm{vec}}\left( {\bf{Z}} \right)$.
Then, we apply $N_{\rm BS}$-point IFFT, i.e., ${\bf F}_{N_{\rm BS}}^{\rm H}$, to the symbols among different antennas for spatial spreading, yielding
\begin{align}
{\bf{S}} = {\bf{ZF}}_{{N_{{\rm{BS}}}}}^{\rm{H}}, \label{signals_after_BF}
\end{align}
where ${\bf{S}}$ is the time-delay-spatial (TDS) domain transmitted symbol matrix of size $MN \times N_{\rm BS}$.
Assuming that a rectangular pulse is applied as the transmitter shaping pulse, it can be shown that the TDS domain transmitted signal for the $n_{\rm t}$-th antenna can be fully characterized by the $n_{\rm t}$-th column of ${\bf{S}}$~\cite{Raviteja2019practical}. Denote by ${\bf s}_{n_{\rm t}}$ the $n_{\rm t}$-th column of ${\bf{S}}$, and we have ${\bf{s}} = {\rm{vec}}\left( {\bf{S}} \right)$, where ${\bf{s}} \buildrel \Delta \over = {\left[ {{\bf{s}}_1^{\rm{H}},{\bf{s}}_2^{\rm{H}},...,{\bf{s}}_{{N_{{\rm{BS}}}}}^{\rm{H}}} \right]^{\rm{H}}}$.
By combining~\eqref{TD_transmitted_symbol_vec},~\eqref{power_allocation_vec_form}, and~\eqref{signals_after_BF}, and considering the property of Kronecker product, we have
\begin{align}
{\bf{s}} = \left( {{\bf{F}}_{{N_{{\rm{BS}}}}}^{\rm{H}} \otimes {{\bf{I}}_{MN}}} \right){\bf{z}} = \left( {{\bf{F}}_{{N_{{\rm{BS}}}}}^{\rm{H}} \otimes {{\bf{I}}_{MN}}} \right)\left( {{\bm{\alpha }} \otimes {{\bf{I}}_{MN}}} \right){\bf{Wv}} = \left( {{\left( {{\bf{F}}_{{N_{{\rm{BS}}}}}^{\rm{H}}{\bm \alpha} } \right)} \otimes {{\bf{I}}_{MN}}} \right){\bf{W}}\left( {{\bf{F}}_N^{\rm{H}} \otimes {{\bf{I}}_M}} \right){\bf{x}},
\label{signals_after_BF_vec}
\end{align}
where ${\bf{W}} \buildrel \Delta \over = {\left[ {{\bf{W}}_1^{\rm{H}},{\bf{W}}_2^{\rm{H}},...,{\bf{W}}_{{N_{{\rm{BS}}}}}^{\rm{H}}} \right]^{\rm{H}}}$ is the concatenated precoding matrix of size $N_{\rm BS}MN \times MN$. For a better understanding, we provide a diagram in Fig.~\ref{domain_transform}, characterizing the domain transformations for the transmitter step by step.
As shown in Fig.~\ref{domain_transform}, the OTFS modulation transforms the broadcast message from the DD domain to the TF domain and then to the TD domain. The precoding and power allocation are performed at the time-delay-angular (TDA) domain after repeating the signals onto each antenna. After that, the IFFT converts the signals from the TDA domain to the  TDS domain for signal transmission.
\subsection{Communication Model}
According to the far-field assumption~\cite{tse2005fundamentals} and the DD domain channel characteristics~\cite{hlawatsch2011wireless},
the communication channel with respect to the antenna index $n_{\rm t}$, for $1 \le n_{\rm t} \le N_{\rm BS}$, and the UE index $i$, for $1 \le i \le K$, can be modeled by
\begin{align}
h\left( {{n_{\rm{t}}},i,\tau ,\nu } \right) = \sum\limits_{p = 1}^P {{h_{i,p}}\exp \left( {j\pi \left( {{n_{\rm{t}}} - 1} \right)\sin {\varphi _{i,p}}} \right)\delta \left( {\tau  - {\tau _{i,p}}} \right)\delta \left( {\nu  - {\nu _{i,p}}} \right)}, \label{com_DD_response}
\end{align}
where we assume that the distance between the adjacent antennas is half of the signal wavelength. In~\eqref{com_DD_response}, $h_{i,p} \in {\mathbb C}$, ${{\varphi _{i,p}}}$, $\tau _{i,p}$, and $\nu _{i,p}$ are the communication fading coefficient, angle of departure (AoD), delay shift, and Doppler shift corresponding to the $p$-th path of the $i$-th UE, respectively.
For the ease of derivation, we assume that the communication fading coefficient $h_{i,p}$ follows the uniform power delay and Doppler profile, such that
the $h_{i,p}$ has zero mean and variance $1/(2P)$ per real dimension for $1 \le p \le P$, and is independent from the delay and Doppler indices~\cite{molisch2012wireless}.
Assuming that a rectangular pulse is applied as the matched-filtering pulse for each UE, then with a reduced cyclic prefix (CP) structure~\cite{Raviteja2019practical}, the TDS domain channel response based on~\eqref{com_DD_response} can be equivalently represented by its matrix form~\cite{Raviteja2019practical}, i.e.,
\begin{align}
{\bf{H}}_{{n_{\rm{t}}},i}^{{\rm{TDS}}} \buildrel \Delta \over = \sum\limits_{p = 1}^P {{h_{i,p}}} \exp \left( {j\pi \left( {{n_{\rm{t}}} - 1} \right)\sin {\varphi _{i,p}}} \right){{\bf{\Pi }}^{{l_{i,p}}}}{{\bf{\Delta }}^{{k_i} + {\kappa _{i,p}}}}, \quad \forall i,p . \label{com_TDS_matrix}
\end{align}
Specifically, we denote by $l_{i,p}$ and $k_{i,p}$ the indices of delay and Doppler associated with the $p$-th path of the $i$-th UE, respectively, where we have
\begin{equation}
{\tau _{i,p}} = \frac{{l_{i,p}}}{{M\Delta f}},\quad {\rm and }\quad
{\nu _{i,p}} = \frac{{{k_{i,p}} + {\kappa _{i,p}}}}{{NT}},
\label{com_DD_resolution}
\end{equation}
respectively. According to the frame format of OTFS modulation, we have $0 \le l_{i,p} \le M-1$, and $0 \le k_{i,p} \le N-1$, respectively~\cite{Raviteja2019practical}.
Note that the term $- {1 \mathord{\left/
 {\vphantom {1 2}} \right.
 \kern-\nulldelimiterspace} 2} \le {\kappa _{i,p}} \le {1 \mathord{\left/
 {\vphantom {1 2}} \right.
 \kern-\nulldelimiterspace} 2}$ denotes the fractional Doppler which corresponds to the fractional shift from the nearest Doppler grid \cite{Raviteja2018interference}. On the other hand, since the typical value of the sampling time ${1 \mathord{\left/
 {\vphantom {1 {M\Delta f}}} \right.
 \kern-\nulldelimiterspace} {M\Delta f}}$ in the delay domain is usually sufficiently small, the impact of
fractional delays in typical wide-band systems can be neglected~\cite{tse2005fundamentals}.
In~\eqref{com_TDS_matrix}, ${\bm{\Pi }}$ is the permutation matrix (forward cyclic shift) characterizing the delay influence, given by
\begin{equation}
{\bm{\Pi }} = {\left[ {\begin{array}{*{20}{c}}
0& \cdots &0&1\\
1& \ddots &0&0\\
 \vdots & \ddots & \ddots & \vdots \\
0& \cdots &1&0
\end{array}} \right]},
\end{equation}
and ${\bm{\Delta}}=\textrm{diag}\{{\gamma}^0,{\gamma}^1,...,{\gamma}^{MN-1}\} $ is a diagonal matrix characterizing the Doppler influence, with ${\gamma} \buildrel \Delta \over = {e^{\frac{{j2\pi }}{{MN}}}}$~\cite{Raviteja2019practical}. With~\eqref{com_TDS_matrix}, the TD domain{\footnote{Since the UE only has one antenna, the spatial/angular domain features from the receiver side, e.g., receiver steering vector, are disappeared. Therefore, we use the term TD instead of TDS/TDA for the relevant descriptions for the communication receiver in sequel. }} received symbol vector for the $i$-th UE is written by
\begin{equation}
{{\bf{r}}_i} = \sum\limits_{{n_{\rm{t}}} = 1}^{{N_{{\rm{BS}}}}} {{\bf{H}}_{{n_{\rm{t}}},i}^{{\rm{TDS}}}{{\bf{s}}_{{n_{\rm{t}}}}} + {{\bf{q }}_i}},  \label{com_TDS_sum_r}
 \end{equation}
where ${{\bf{q }}_i}$ denotes the additive white Gaussian noise (AWGN) samples with one-sided power spectral density (PSD) $N_0$.
Equivalently, by separating the angular features in~\eqref{com_TDS_matrix},~\eqref{com_TDS_sum_r} can be rearranged as
\begin{align}
{{\bf{r}}_i} = \sum\limits_{p = 1}^{P} \left( {{{\bf{a}}^{\rm{T}}}\left( {{\varphi _{i,p}}} \right) \otimes {\bf{H}}_{i,p}^{{\rm{TD}}}} \right){\bf{s}} + {{\bf{q }}_i},
\end{align}
where ${\bf{a}}\left( {{\varphi _{i,p}}} \right)$ is the transmit steering vector given by
\begin{align}
{\bf{a}}\left( {{\varphi _{i,p}}} \right) \buildrel \Delta \over = \frac{1}{{\sqrt {{N_{{\rm{BS}}}}} }}{\left[ {1,\exp \left( {j\pi \sin {\varphi _{i,p}}} \right),...,\exp \left( {j\pi \left( {{N_{{\rm{BS}}}} - 1} \right)\sin {\varphi _{i,p}}} \right)} \right]^{\rm{T}}},
\label{steering_vector}
\end{align}
and
${{\bf{H}}_{i,p}^{{\rm{TD}}}}$ is defined as the TD domain equivalent communication channel matrix for the $p$-th path of the $i$-th UE given by
\begin{align}
{\bf{H}}_{i,p}^{{\rm{TDP}}} \buildrel \Delta \over =  {{h_{i,p}}} {{\bf{\Pi }}^{{l_{i,p}}}}{{\bf{\Delta }}^{{k_{i,p}} + {\kappa _{i,p}}}} . \label{com_TD_P_matrix}
\end{align}
Then, according to the connections between the TD domain to the DD domain~\cite{Raviteja2019practical}, the DD domain received signal for the $i$-th UE is given by
\begin{align}
{{\bf{y}}_i} = \left( {{{\bf{F}}_N} \otimes {{\bf{I}}_M}} \right)\sum\limits_{p = 1}^P {\left( {{{\bf{a}}^{\rm{T}}}\left( {{\varphi _{i,p}}} \right) \otimes {\bf{H}}_{i,p}^{{\rm{TD}}}} \right)} {\bf{s}} + {{\bm{\eta }}_i}, \label{com_DD_y-der1}
\end{align}
where ${{\bm{\eta }}_i}  \buildrel \Delta \over =  \left( {{{\bf{F}}_N} \otimes {{\bf{I}}_M}} \right){{\bf{q}}_i}$ is the equivalent AWGN noise vector in the DD domain.
Finally, by substituting~\eqref{signals_after_BF_vec} into~\eqref{com_DD_y-der1}, we obtain
\begin{align}
{{\bf{y}}_i} &= \left( {{{\bf{F}}_N} \otimes {{\bf{I}}_M}} \right)\sum\limits_{p = 1}^P {\left( {{{\bf{a}}^{\rm{T}}}\left( {{\varphi _{i,p}}} \right) \otimes {\bf{H}}_{i,p}^{{\rm{TD}}}} \right)} \left( {{\left( {{\bf{F}}_{{N_{{\rm{BS}}}}}^{\rm{H}}{\bm \alpha} } \right)} \otimes {{\bf{I}}_{MN}}} \right){\bf{W}}\left( {{\bf{F}}_N^{\rm{H}} \otimes {{\bf{I}}_M}} \right){\bf{x}} + {{\bm{\eta }}_i}\notag\\
& = \left( {{{\bf{F}}_N} \otimes {{\bf{I}}_M}} \right)\sum\limits_{p = 1}^P {\left( {\left( {{{\bf{a}}^{\rm{T}}}\left( {{\varphi _{i,p}}} \right){\bf{F}}_{{N_{{\rm{BS}}}}}^{\rm{H}}{\bm \alpha}} \right) \otimes {\bf{H}}_{i,p}^{{\rm{TD}}}} \right)} {\bf{W}}\left( {{\bf{F}}_N^{\rm{H}} \otimes {{\bf{I}}_M}} \right){\bf{x}} + {{\bm{\eta }}_i}. \label{com_DD_y-der2}
\end{align}
\subsection{Radar Model}
Similar to the communication model, we consider the radar channel response with respect to the transmit antenna index $n_{\rm t}$, for $1 \le n_{\rm t} \le N_{\rm BS}$, the UE index $i$, for $1 \le i \le K$, the receive antenna index $n_{\rm r}$, for $1 \le n_{\rm r} \le N_{\rm BS}$, which is modeled by
\begin{align}
&\tilde h\left( {{n_{\rm{t}}},i,{n_{\rm{r}}},\tilde \tau ,\tilde \nu } \right) \notag\\
=& \sum\limits_{p = 1}^P {{{\tilde h}_{i,p}}\exp \left( {j\pi \left( {{n_{\rm{t}}} - 1} \right)\sin {\varphi _{i,p}}} \right)\exp \left( {j\pi \left( {{n_{\rm{r}}} - 1} \right)\sin {\varphi _{i,p}}} \right)\delta \left( {\tau  - {{\tilde \tau }_{i,p}}} \right)\delta \left( {\nu  - {{\tilde \nu }_{i,p}}} \right)} , \label{radar_DD_response}
\end{align}
where we assume that the receive and transmit antennas are co-located such that AoDs and angle of arrival (AoAs) are of the same values.
In~\eqref{radar_DD_response}, ${\tilde h}_{i,p}\forall {\mathbb C}$, ${\tilde \tau} _{i,p}$, and ${\tilde \nu} _{i,p}$ are the radar reflection coefficient, round-trip delay shift, and round-trip Doppler shift associated to the $p$-th path of the $i$-th UE, respectively, where the round-trip delay and Doppler shifts satisfy ${{{\tilde \tau }_{i,p}}}=2{{{ \tau }_{i,p}}}$ and ${{{\tilde \nu }_{i,p}}}=2{{{ \nu }_{i,p}}}$~\cite{liu2020joint,yuan2021integrated}.
In practice, the radar reflection coefficient ${\tilde h}_{i,p}$ relates to the distance between the $i$-th UE and the BS with respect to the $p$-th path, the effective area of the radar receiving antenna, the RCS, the wave length, and the transmit and receive antenna gains~\cite{shrestha2008method}. 
Similar to the communication model, we assume that a rectangular pulse is applied as the filtering pulse at the receiver, then with the reduced CP structure~\cite{Raviteja2019practical}, the TDS domain equivalent radar sensing matrix is given by~\cite{Raviteja2019practical}
\begin{align}
{\bf{\tilde H}}_{{n_{\rm{t}}},i,{n_{\rm{r}}}}^{{\rm{TDS}}} \buildrel \Delta \over = \sum\limits_{p = 1}^P {{{\tilde h}_{i,p}}} \exp \left( {j\pi \left( {{n_{\rm{t}}} - 1} \right)\sin {\varphi _{i,p}}} \right)\exp \left( {j\pi \left( {{n_{\rm{r}}} - 1} \right)\sin {\varphi _{i,p}}} \right){{\bf{\Pi }}^{{{\tilde l}_{i,p}}}}{{\bf{\Delta }}^{{{\tilde k}_i} + {{\tilde \kappa }_{i,p}}}}, \label{radar_TDS_matrix}
\end{align}
where ${{{\tilde l}_{i,p}}}=2{{{ l}_{i,p}}}$ and ${{\tilde k}_i} + {{\tilde \kappa }_{i,p}} = 2\left( {{k_i} + {\kappa _{i,p}}} \right)$.
Let ${{\tilde N}_0}$ denote the one-sided PSD for the radar noise, which takes into account of both the AWGN noise power from the channel and the interference power from the transmit signals after interference cancellation~\cite{liu2020joint}. Then, similar to~\eqref{com_TDS_sum_r}, the radar received TDS domain symbol vector for the $n_{\rm r}$-th antenna is written by
\begin{equation}
{{{\bf{\tilde r}}}_{{n_{\rm{r}}}}} = \sum\limits_{i = 1}^K {\sum\limits_{{n_{\rm{t}}} = 1}^{{N_{{\rm{BS}}}}} {{\bf{\tilde H}}_{{n_{\rm{t}}},i,{n_{\rm{r}}}}^{{\rm{TDS}}}{{\bf{s}}_{{n_{\rm{t}}}}} + {{{\bf{\tilde q}}}_{n_{\rm r}}}} }  ,  \label{radar_TDS_sum_r}
 \end{equation}
where ${{{\bf{\tilde q}}}_{n_{\rm r}}} $ denotes the TDS domain radar AWGN vector.
By stacking the TDS domain received symbols from each antenna, we have
\begin{align}
{\bf{\tilde r}} = {\left[ {{\bf{\tilde r}}_1^{\rm{H}},{\bf{\tilde r}}_2^{\rm{H}},...,{\bf{\tilde r}}_{{N_{{\rm{BS}}}}}^{\rm{H}}} \right]^{\rm{H}}}
= \sum\limits_{i = 1}^K {\sum\limits_{p = 1}^P {\left( {{\bf{A}}\left( {{\varphi _{i,p}}} \right) \otimes {\bf{\tilde H}}_{i,p}^{{\rm{TD}}}} \right)} } {\bf{s}} + {\bf{\tilde q}}, \label{radar_TDS_vec_r_der1}
\end{align}
where ${\bf{\tilde q}}  \buildrel \Delta \over =  {\left[ {{\bf{\tilde q}}_1^{\rm{H}},{\bf{\tilde q}}_2^{\rm{H}},...,{\bf{\tilde q}}_{{N_{{\rm{BS}}}}}^{\rm{H}}} \right]^{\rm{H}}}$ is the equivalent TDS domain radar noise,
\begin{align}
{\bf{A}}\left( {{\varphi _{i,p}}} \right) \buildrel \Delta \over = {\bf{a}}\left( {{\varphi _{i,p}}} \right){{\bf{a}}^{\rm{T}}}\left( {{\varphi _{i,p}}} \right), \forall i, p, \label{radar_steering_mtx}
\end{align}
is the steering matrix associated with the $p$-th path of the $i$-th UE, and
\begin{align}
{\bf{\tilde H}}_{i,p}^{{\rm{TD}}} \buildrel \Delta \over = {{{\tilde h}_{i,p}}} {{\bf{\Pi }}^{{{\tilde l}_{i,p}}}}{{\bf{\Delta }}^{{{\tilde k}_i} + {{\tilde \kappa }_{i,p}}}}.\label{radar_TD_P_matrix}
\end{align}
is the TDS domain equivalent radar channel matrix for the $p$-th path of the $i$-th UE, respectively.
Without loss of generality, we consider the radar sensing in the TDA domain. Therefore, we apply the spatial de-spreading to the TDS domain radar received symbol vector ${\bf{\tilde r}}$, yielding
\begin{align}
{\bf{\tilde z}} &= \left( {{{\bf{F}}_{{N_{{\rm{BS}}}}}} \otimes {{\bf{I}}_{MN}}} \right){\bf{\tilde r}} = \left( {{{\bf{F}}_{{N_{{\rm{BS}}}}}} \otimes {{\bf{I}}_{MN}}} \right)\sum\limits_{i = 1}^K {\sum\limits_{p = 1}^P {\left( {{\bf{A}}\left( {{\varphi _{i,p}}} \right) \otimes {\bf{\tilde H}}_{i,p}^{{\rm{TD}}}} \right)\left( {{\bf{F}}_{{N_{{\rm{BS}}}}}^{\rm{H}} \otimes {{\bf{I}}_{MN}}} \right){\bf{z}}}  + {\bm{\tilde \eta}}} \notag\\
& = \sum\limits_{i = 1}^K {\sum\limits_{p = 1}^P {\left( {\left( {{{\bf{F}}_{{N_{{\rm{BS}}}}}}{\bf{A}}\left( {{\varphi _{i,p}}} \right){\bf{F}}_{{N_{{\rm{BS}}}}}^{\rm{H}}{\bm{\alpha }}} \right) \otimes {\bf{\tilde H}}_{i,p}^{{\rm{TD}}}} \right){\bf{d}}}  + {\bm{\tilde \eta}} }  , \label{radar_TDA_after_de_spreading}
\end{align}
where ${\bm{\tilde \eta}}$ is the TDA domain equivalent radar AWGN vector with one-sided PSD ${{\tilde N}_0}$.

\subsection{Model Simplifications with Spatial Spreading and De-spreading}
To characterize the effect of spatial spreading and de-spreading, we are interested in the structure of the equivalent angular domain channel vector/matrix for both communication and radar.
For the $p$-th path of the $i$-th UE, let us define the equivalent angular domain channel vector for the communication channel by
\begin{align}
{\bf{h}}_{i,p}^{\rm{A}} \buildrel \Delta \over = {{\bf{a}}^{\rm{T}}}\left( {{\varphi _{i,p}}} \right){\bf{F}}_{{N_{{\rm{BS}}}}}^{\rm{H}}{\bm \alpha}, \label{com_H_A}
\end{align}
and the equivalent angular domain channel matrix for the radar channel by
\begin{align}
{\bf{\tilde H}}_{i,p}^{\rm{A}} \buildrel \Delta \over = {{\bf{F}}_{{N_{{\rm{BS}}}}}}{\bf{A}}\left( {{\varphi _{i,p}}} \right){\bf{F}}_{{N_{{\rm{BS}}}}}^{\rm{H}}{\bm \alpha},\label{radar_H_A}
\end{align}
respectively.
Based on~\eqref{com_H_A} and~\eqref{radar_H_A}, we can derive the elements in ${\bf{h}}_{i,p}^{\rm{A}}$ and ${\bf{\tilde H}}_{i,p}^{\rm{A}}$ after some manipulations.
In particular, for $1 \le k \le N_{\rm BS}$, and $1 \le l \le N_{\rm BS}$, we have
\begin{align}
{{h}}_{i,p}^{\rm{A}}\left[ l \right]= \frac{\sqrt{\alpha_l}}{{{N_{{\rm{BS}}}}}}\left( {\frac{{1 - \exp \left( {j\pi {N_{{\rm{BS}}}}\sin {\varphi _{i,p}}} \right)}}{{1 - \exp \left( {j\pi \sin {\varphi _{i,p}} + j2\pi \frac{l-1}{{{N_{{\rm{BS}}}}}}} \right)}}} \right), \label{com_H_A_value}
\end{align}
and
\begin{align}
{{\tilde H}}_{i,p}^{\rm{A}}\left[ k,l \right]=  \frac{\sqrt{\alpha_l}}{{{{\left( {{N_{{\rm{BS}}}}} \right)}^2}}}\left( {\frac{{1 - \exp \left( {j\pi {N_{{\rm{BS}}}}\sin {\varphi _{i,p}}} \right)}}{{1 - \exp \left( {j\pi \sin {\varphi _{i,p}} - j2\pi \frac{k-1}{{{N_{{\rm{BS}}}}}}} \right)}}} \right)\left( {\frac{{1 - \exp \left( {j\pi {N_{{\rm{BS}}}}\sin {\varphi _{i,p}}} \right)}}{{1 - \exp \left( {j\pi \sin {\varphi _{i,p}} + j2\pi \frac{l-1}{{{N_{{\rm{BS}}}}}}} \right)}}} \right). \label{radar_H_A_value}
\end{align}
Based on~\eqref{com_H_A_value} and~\eqref{radar_H_A_value}, we notice that the if the value of $\sin {\varphi _{i,p}}$ is multiples of $2/N_{\rm BS}$, the values of ${{h}}_{i,p}^{\rm{A}}\left[ l \right]$ and ${{\tilde H}}_{i,p}^{\rm{A}}\left[ k,l \right]$ will be either $\sqrt {\alpha_l}$ or zero.
Therefore, let us define the \textbf{angular resolution} by $2/N_{\rm BS}$.
According to the angular resolution, we further define the \textbf{transmit angular index} by ${a_{i,p}} \buildrel \Delta \over = {\left[ {{N_{{\rm{BS}}}} - \frac{{\sin \left( {{\varphi _{i,p}}} \right){N_{{\rm{BS}}}}}}{2}} \right]_{{N_{{\rm{BS}}}}}}+1$ and the \textbf{receive angular index} by ${{\tilde a}_{i,p}} \buildrel \Delta \over = {\left[ {{N_{{\rm{BS}}}} + \frac{{\sin \left( {{\varphi _{i,p}}} \right){N_{{\rm{BS}}}}}}{2}} \right]_{{N_{{\rm{BS}}}}}}+1$.
Based on these, we have the following lemma.

\textbf{Lemma 1} (\emph{Asymptotical Orthogonality}):
With a sufficiently large number of antennas at the BS, the angular index ${a_{i,p}} $ is of an integer value. In this case,
we have ${{h}}_{i,p}^{\rm{A}}\left[ l \right]=0$, for $ l \ne {{a_{i,p}}}$, while ${{h}}_{i,p}^{\rm{A}}\left[ l \right]=\sqrt{\alpha_l}$, for $ l = {{a_{i,p}}}$. Furthermore, we have ${{\tilde H}}_{i,p}^{\rm{A}}\left[ k,l \right]=0$, for $ k \ne {\tilde a}_{i,p} , l \ne {{a_{i,p}}}$, while ${{\tilde H}}_{i,p}^{\rm{A}}\left[ k,l \right]=\sqrt{\alpha_l}$, for $ k ={{\tilde a}_{i,p}} , l = {a_{i,p}}$.

\emph{Proof}: The proof is derived based on the straightforward calculations of~\eqref{com_H_A_value} and~\eqref{radar_H_A_value}. Hence, it is omitted due to the space limitation.

\begin{figure}[htbp]
\centering
\subfigure[Absolute values of ${\bf{\tilde H}}_{i,p}^{\rm{A}}$ without spreading and de-spreading.]{
\begin{minipage}[t]{0.5\textwidth}
\centering
\includegraphics[scale=0.5]{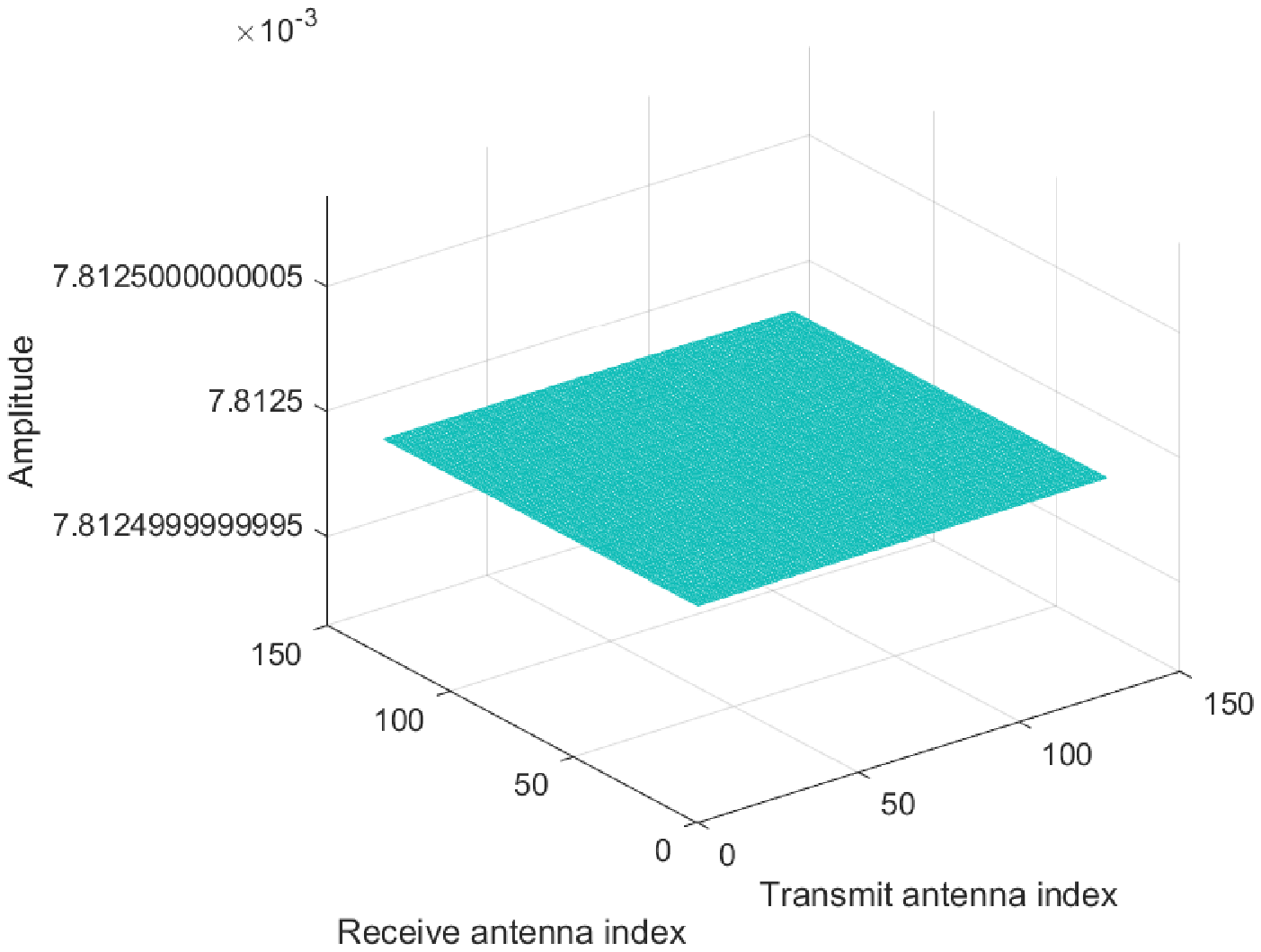}
\label{withoutSS}
\end{minipage}%
}%
\subfigure[Absolute values of ${\bf {\tilde H}}_{i,p}^{\rm{A}}$ with spreading and de-spreading.]{
\begin{minipage}[t]{0.5\textwidth}
\centering
\includegraphics[scale=0.5]{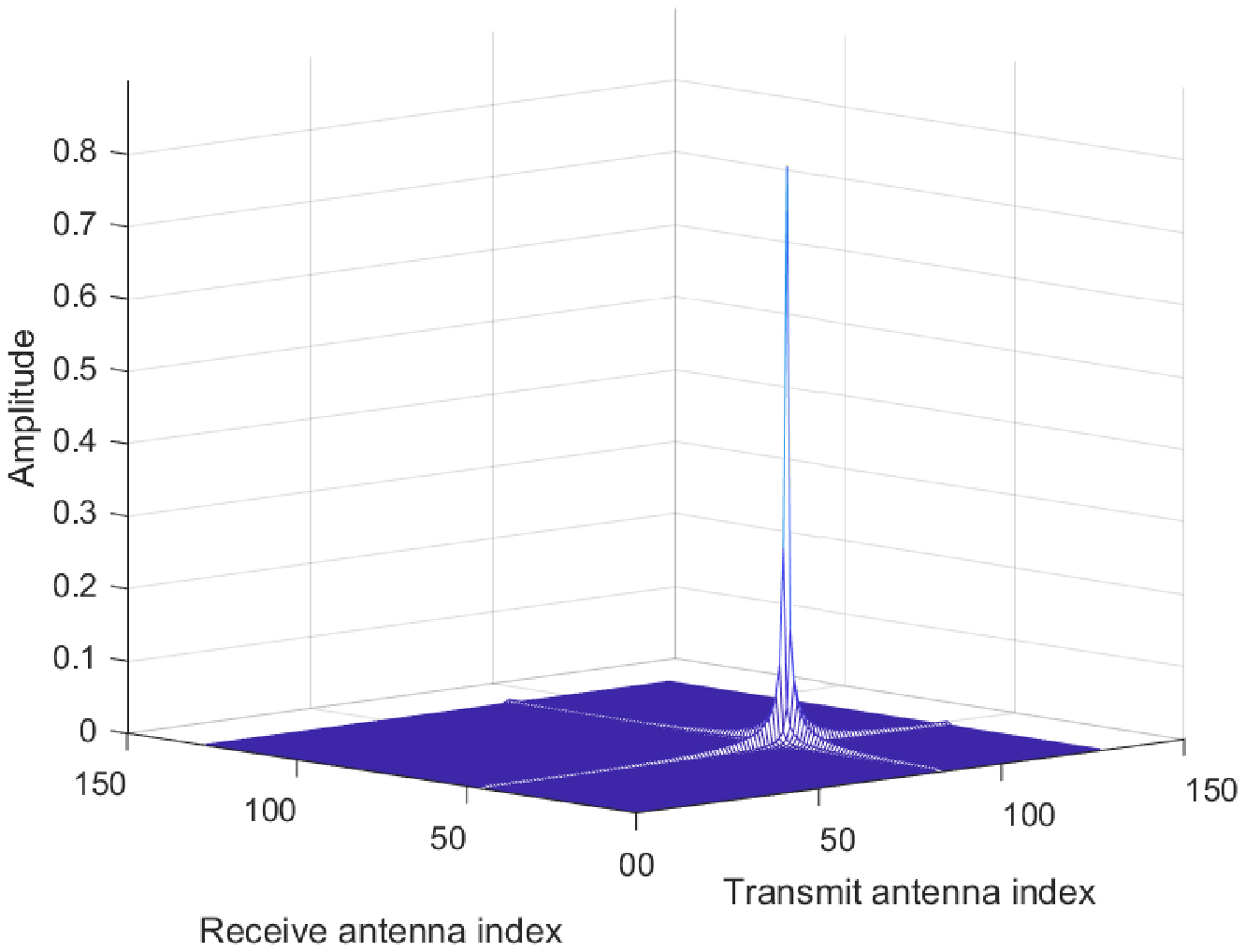}
\label{withSS}
\end{minipage}%
}%
\centering
\caption{Equivalent angular domain channel matrix for radar, where the AoA is $\varphi _{i,p}=\pi/4$ and $N_{\rm BS}=128$.}
\vspace{-5mm}
\label{withou_and_without_SS}
\end{figure}

As indicated from Lemma 1, we notice that the interference for both the communication and radar channels due to the multipath and multiuser transmissions can be approximately eliminated by spatial spreading and de-spreading, given a sufficient number of antennas.
For a better understanding, we provide an example of the equivalent angular domain radar channel matrix with $\varphi _{i,p}=\pi/4$ and $N_{\rm BS}=128$ in Fig.~\ref{withou_and_without_SS}. As indicated by the figure, the channel matrix with spatial spreading and de-spreading is much more sparse, i.e., only the grids around the $84$-th transmitted antenna and the $46$-th received antenna are of values, which is consistent with our analysis.
In what follows, we assume that $N_{\rm BS}$ is sufficiently large such that the angular indices are of integer values{\footnote{Given the accuracy of AoD and AoA estimates for practical systems, this assumption is valid for practical BS setups of ISAC transmissions~\cite{liu2020joint,yuan2021integrated}. }}.
Particularly, notice that when $N_{\rm BS}$ is sufficiently large, the angular resolution is sufficiently high. Hence, we further assume that all the paths can be fully separated by its angular domain features, i.e., $ {{a_{i,p}} \ne {a_{i',p'}}}$, for $i \ne i'$ or $p \ne p'$.
Consequently, both the communication and radar models in~\eqref{com_DD_y-der2} and~\eqref{radar_TDA_after_de_spreading} can be further simplified, yielding
\begin{align}
{{\bf{y}}_i} = \sum\limits_{p = 1}^P {\sqrt{\alpha_{a_{i,p}}}}{\left( {{{\bf{F}}_N} \otimes {{\bf{I}}_M}} \right){\bf{H}}_{i,p}^{{\rm{TD}}}{{\bf{W}}_{{a_{i,p}}}}\left( {{\bf{F}}_N^{\rm{H}} \otimes {{\bf{I}}_M}} \right){\bf{x}} + {{\bm{\eta }}_i}} ,\label{com_model}
\end{align}
and
\begin{align}
{\bf{\tilde z}} = {\bf{\tilde H}}_{{\rm{Radar}}}^{{\rm{TDA}}}{\bf{z}} + {\bm{\tilde \eta}}  ,\label{radar_model}
\end{align}
respectively, where ${\bf{\tilde H}}_{{\rm{Radar}}}^{{\rm{TDA}}}$ is the TDA domain effective radar sensing matrix.
In particular, it can be shown that ${\bf{\tilde H}}_{{\rm{Radar}}}^{{\rm{TDA}}}$
is a block matrix of size $N_{\rm BS}MN \times N_{\rm BS}MN$, whose $({\tilde a}_{i,p}, a_{i,p})$-th sub-block is given by ${\sqrt{\alpha_{a_{i,p}}}}{\bf{\tilde H}}_{i,p}^{{\rm{TD}}}$ and the rest sub-blocks are given by ${{\bf{0}}_{MN}}$.

\textbf{Remark 1}: As indicated by the above discussion, the proposed spatial spreading and de-spreading leads to the discretization of angular features, which simplifies the input-output relationship for both communication and radar channels. In particular, the angular domain discretization enables direct interactions between the transmitted signal on a specific antenna and the channel distortion associated to a specific path.
Therefore, it is suitable for ISAC transmissions.
\section{Radar Sensing Designs based on SS-OTFS Modulation}
In this section, we are focusing on the radar sensing designs based on the SS-OTFS modulation. According to the angular discretization enabled by spatial spreading and de-spreading, we aim to design practical beam tracking and AoA estimation algorithms based on the previous AoA estimates and radar reflection coefficients. It should be noted that, from the radar point of view, not all the targets are of interest, because they may introduce clutter interference that can potentially undermine the sensing performance. However, some of the clutter might come from scatterers that can significantly contribute to the total received power for communication transmissions. Therefore, it still be necessary to estimate the parameters corresponding all the scatterers
for ISAC systems~\cite{liu2020joint}.

\subsection{Beam Tracking}
Let us focus on the radar channel model given in~\eqref{radar_model}. In particular, with a sufficiently large number of antennas, the effective radar sensing matrix ${\bf{\tilde H}}_{{\rm{Radar}}}^{{\rm{TDA}}}$ is a sparse block matrix as discussed in Section II. Thanks to the sparsity enabled by the SS-OTFS transmission, the angular domain is sufficiently discretized. Therefore, the signals can be transmitted towards the desired directions by simply assigning power to the corresponding antennas without sophisticated precoding designs.

For the case of beam tracking, the BS knows the previous AoA estimates associated to the different paths of the UEs from the previous time instant.
Therefore, a common approach for beam tracking is to send relatively wider beams towards the AoAs estimated from the previous time instant.
To achieve this, we only need to allocate power on the corresponding antennas. Denote by ${\theta _{{\rm{range}}}}$ the desired beam width. According to the angular resolution, we need to allocate powers on ${N_{{\rm{range}}}} \buildrel \Delta \over = \frac{{{\theta _{{\rm{range}}}}}}{2}{N_{{\rm{BS}}}}$ antennas. In specific, let ${{\mathbb U}_{i,p}}$ denote the antenna set corresponding to those antennas for the $p$-th paths of the $i$-th UE and it is given by
\begin{align}
{{\mathbb U}_{i,p}}  \buildrel \Delta \over =  \left[ {{{\left[ {{N_{{a_{i,p}}}} \!- \! 1 \!-\! \frac{{{N_{{\rm{range}}}}}}{2}} \right]}_{{N_{{\rm{BS}}}}}}\! \!\!+ \!1,{{\left[ {{N_{{a_{i,p}}}} \!- \!1 \!- \frac{{{N_{{\rm{range}}}}}}{2}} \right]}_{{N_{{\rm{BS}}}}}}\!\!\! + \!1,...,{{\left[ {{N_{{a_{i,p}}}}\!-1 \!+ \frac{{{N_{{\rm{range}}}}}}{2}} \right]}_{{N_{{\rm{BS}}}}}}\! \!\!+ \!1} \right],
\label{antenna_set}
\end{align}
where we assume that ${N_{{\rm{range}}}}$ is an even number.
Given the high angular resolution, we further assume that the AoAs are sufficiently separated, such that ${{\mathbb U}_{i,p}} \cap {{\mathbb U}_{i',p'}} = \varnothing , \forall i \ne i',p \ne p'$.
To ensure the reliability of radar sensing and communication, we propose to send the precoded signals with equal powers towards all the antennas in ${{\mathbb U}_{i,p}}$, such that the ISAC signal will be guaranteed to be reflected/received by the corresponding UE.


\subsection{AoA Estimation}
Now let us focus on the AoA estimation.
According to the radar model in~\eqref{radar_model}, we notice that the TDA domain equivalent radar sensing matrix is a sparse block matrix, where only the sub-blocks related to the angular indices ${\tilde a}_{i,p}$ are of non-zero values. Noticing that the number of non-zero sub-blocks in any row/column partitions of ${\bf{\tilde H}}_{{\rm{Radar}}}^{{\rm{TDA}}}$ cannot be larger than one, the ${{\tilde a}_{i,p}}$-th diagonal sub-block in the covariance matrix of ${\bf{\tilde z}}$ is written by
\begin{align}
{{\bf{R}}_{{\bf{\tilde z}}}}\left[ {{{\tilde a}_{i,p}},{{\tilde a}_{i,p}}} \right] = {\alpha _{{{ a}_{i,p}}}}{\bf{\tilde H}}_{i,p}^{{\rm{TD}}}{{\bf{W}}_{{a_{i,p}}}}{\bf{W}}_{{a_{i,p}}}^{\rm{H}}{\left( {{\bf{\tilde H}}_{i,p}^{{\rm{TD}}}} \right)^{\rm{H}}} + {\tilde N_0}{{\bf{I}}_{MN}}, \label{AoA_der1}
\end{align}
where we assume that ${\bf{x}}{{\bf{x}}^{\rm{H}}} = {{\bf{I}}_{MN}}$.
Furthermore, by noticing that ${{\bf{W}}_{{{ a}_{i,p}}}}{\bf{W}}_{{{ a}_{i,p}}}^{\rm{H}} = {{\bf{I}}_{MN}}$ according to our definition,~\eqref{AoA_der1} can be further simplified by
\begin{align}
{{\bf{R}}_{{\bf{\tilde z}}}}\left[ {{{\tilde a}_{i,p}},{{\tilde a}_{i,p}}} \right] = \left( {{\alpha _{{{ a}_{i,p}}}}{{\left| {{{\tilde h}_{i,p}}} \right|}^2} + {{\tilde N}_0}} \right){{\bf{I}}_{MN}}. \label{AoA_der2}
\end{align}
Based on~\eqref{AoA_der2}, a straightforward AoA estimation design is to find the indices associated to the $KP$ largest values of the traces of the diagonal sub-blocks and then derive the corresponding AoAs associated to the receive antenna indices ${{{\tilde a}_{i,p}}}$.
An important issue at this point is to associate the estimated AoAs with the paths of different UEs. Unfortunately, we do not have enough space to discuss this issue. We refer the interested readers to~\cite{liu2020joint} for more information.
\subsection{Power Allocation for Radar Sensing}
As indicated by~\eqref{AoA_der2}, the power allocation will affect estimation performance. In particular, the value of ${{\left| {{{\tilde h}_{i,p}}} \right|}^2}$ can be largely different for different paths. Therefore, it is important to assign different powers to the related antennas according to the value of ${{\left| {{{\tilde h}_{i,p}}} \right|}^2}$.
For the ease of derivation, let us assume that the value of ${{\left| {{{\tilde h}_{i,p}}} \right|}^2}$ of the current time instant is accurately derived based on the corresponding estimates from the previous time instant. As the above AoA estimation is based on the matched-filtering principle, we aim to maximize the minimum effective radar SNR among all the paths, such that the radar can provide a relatively accurate sensing performance for each path.
Notice that ${\tilde N}_0$ is assumed to be equal to each UE. Therefore, to maximize the minimum radar SNR is equivalent to maximize the received power ${{\alpha _{{a_{i,p}}}}{{\left| {{{\tilde h}_{i,p}}} \right|}^2}}$ associated to each path.
Therefore, our power allocation is designed as follows
\begin{align}
\mathop {\max }\limits_{{\alpha _{{a_{1,1}}}},...,{\alpha _{{a_{K,P}}}}}&\quad {\left( {{\alpha _{{a_{i,p}}}}{{\left| {{{\tilde h}_{i,p}}} \right|}^2}} \right)_{\min }} \label{maximum_power}\\
{\rm{s}}{\rm{.t}}{\rm{.}}&\quad\left( {\sum\limits_{i = 1}^K {\sum\limits_{p = 1}^P {{\alpha _{{a_{i,p}}}}} } } \right)\left( {{N_{{\rm{range}}}} + 1} \right) = {\alpha _{{\rm{total}}}}\label{sum_power_constraint}
\end{align}
It can be shown that the maximum of~\eqref{maximum_power} is achieved when all the received power are of the same value, i.e.,
${\alpha _{{a_{1,1}}}}{\left| {{{\tilde h}_{1,1}}} \right|^2} = {\alpha _{{a_{1,2}}}}{\left| {{{\tilde h}_{1,2}}} \right|^2} = ... = {\alpha _{{a_{K,P}}}}{\left| {{{\tilde h}_{K,P}}} \right|^2}$.
Therefore, according to~\eqref{sum_power_constraint}, the power allocation should satisfy
\begin{align}
{\alpha _{{a_{i,p}}}} = {{\left( {\frac{{{\alpha _{{\rm{total}}}}}}{{{N_{{\rm{range}}}} + 1}}\frac{1}{{{{\left| {{{\tilde h}_{i,p}}} \right|}^2}}}} \right)} \mathord{\left/
 {\vphantom {{\left( {\frac{{{\alpha _{{\rm{total}}}}}}{{{N_{{\rm{range}}}} + 1}}\frac{1}{{{{\left| {{{\tilde h}_{i,p}}} \right|}^2}}}} \right)} {\left( {\sum\limits_{j = 1}^K {\sum\limits_{p' = 1}^P {\frac{1}{{{{\left| {{{\tilde h}_{j,p'}}} \right|}^2}}}} } } \right)}}} \right.
 \kern-\nulldelimiterspace} {\left( {\sum\limits_{j = 1}^K {\sum\limits_{p' = 1}^P {\frac{1}{{{{\left| {{{\tilde h}_{j,p'}}} \right|}^2}}}} } } \right)}}.
\end{align}

\textbf{Remark 2}: Intuitively speaking, the proposed power allocation aims to assign a larger power towards to the direction where the power of the radar echo is small. This is contradict to the conventional water-filling principle from the communication point of view, where more power should be assigned towards the direction with a larger channel gain in order to improve the achievable rate. However, since the communication fading coefficients cannot be directly obtained based on the radar sensing estimates, it may be difficult for the BS to design a suitable power allocation that can provide a good trade-off for both radar sensing and communication performances. Consequently, the best we can do is to adapt the proposed power allocation with respect to the statistical distribution of the communication fading coefficients. The related issues will be discussed in detail in Section IV-C.

\section{Sensing-Assisted Communication Design}
In this section, we will develop communication schemes based on the estimated parameters from radar sensing. We notice that there is a deterministic relationship between the round-trip delay and Doppler from radar sensing and the delay and Doppler for communication. Furthermore, given the AoA estimates from radar, we can also determine the corresponding transmit angular indices ${a_{i,p}}$. Unfortunately, there is no direct relationship between the communication fading coefficients and radar reflection coefficients. Therefore, our design criterion is to minimize the PEP for communication with respect to the \emph{a priori} AoA, delay and Doppler estimates that are obtained from radar sensing.
According to the quasi-static property of the DD domain channel response~\cite{Zhiqiang_magzine,Li2020on,hlawatsch2011wireless}, we assume that the minor changes of the related parameters from the previous time instant are
well-compensated~\cite{liu2020joint}, including the delay shifts, and Doppler shifts associated to the corresponding paths.
On the other hand, we note that the AoAs at the current time instant may be different from the estimates from the previous time instant, due to the high angular resolution and mobility. However, as will be explained in detail later, our proposed design can be easily combined with the beam tracking scheme introduced in the previous section. Therefore, we first assume that the AoA estimates are accurate in this section for the ease of derivation.

\subsection{Pair-wise Error Probability Analysis}
Recalling~\eqref{com_model}, we define the effective DD domain communication channel matrix for the $i$-th UE by
\begin{align}
{\bf{H}}_i^{{\rm{DD}}} \buildrel \Delta \over = \sum\limits_{p = 1}^P {{\sqrt{\alpha_{a_{i,p}}}}{h_{i,p}}} \left( {{{\bf{F}}_N} \otimes {{\bf{I}}_M}} \right){{\bf{\Pi }}^{{l_{i,p}}}}{{\bf{\Delta }}^{{k_{i,p}} + {\kappa _{i,p}}}}{{\bf{W}}_{{a_{i,p}}}}\left( {{\bf{F}}_N^{\rm{H}} \otimes {{\bf{I}}_M}} \right).
\label{H_DD_i}
\end{align}
Then,~\eqref{com_model} can be rewritten by
\begin{align}
{{\bf{y}}_i} = {\bf{H}}_i^{{\rm{DD}}}{\bf{x}} + {{\bm \eta}_i}.
\label{new_com_model}
\end{align}
In what follows, we will study the PEP performance with the ML detection in order to facilitate our precoding design.
For the ease of presentation, let us define the following vectors and matrices for related parameters. We define the
effective fading coefficients by ${\bf{h}}_i^{{\rm{eff}}} \buildrel \Delta \over = {\left[ { {h_{i,1}}, {h_{i,2}},..., {h_{i,P}}} \right]^{\rm{T}}}$, the delay shifts by
${\bm{\omega }}_i^\tau  \buildrel \Delta \over = {\left[ {{l_{i,1}},{l_{i,2}},...,{l_{i,P}}} \right]^{\rm{T}}}$, the Doppler shifts by
${\bm{\omega }}_i^\nu  \buildrel \Delta \over = {\left[ {{k_{i,1}} + {\kappa _{i,1}},{k_{i,2}} + {\kappa _{i,2}},...,{k_{i,P}} + {\kappa _{i,P}}} \right]^{\rm{T}}}$, the precoding matrices by ${\bf{W}}_i^{{\rm{com}}} \buildrel \Delta \over = {\left[ {{\bf{W}}_{{a_{i,1}}}^{\rm{H}},{\bf{W}}_{{a_{i,2}}}^{\rm{H}},...,{\bf{W}}_{{a_{i,P}}}^{\rm{H}}} \right]^{\rm{H}}}$, and the allocated power for the $i$-th UE by
${\bm \alpha} _i^{{\rm{com}}} \buildrel \Delta \over = {\rm{diag}}\left\{ {\sqrt {{\alpha _{{a_{i,1}}}}} ,\sqrt {{\alpha _{{a_{i,2}}}}} ,...,\sqrt {{\alpha _{{a_{i,P}}}}} } \right\}$, respectively.
According to \cite{Raviteja2019effective} and \cite{li2021performance},~\eqref{new_com_model} can be rewritten by
\begin{align}
{{\bf{y}}_i} = {\bf{\Phi }}_i^{\omega _i^\tau ,\omega _i^{_\nu },{\bf{W}}_i^{{\rm{com}}}}\left( {\bf{x}} \right){\bm \alpha} _i^{{\rm{com}}}{\bf{h}}_i^{{\rm{eff}}} + {{\bm \eta} _i} \label{new_com_model2},
\end{align}
where ${\bf{\Phi }}_i^{\omega _i^\tau ,\omega _i^{_\nu },{\bf{W}}_i^{{\rm{com}}}}\left( {\bf{x}} \right)$ is referred to as the \emph{equivalent codeword matrix} and it is a concatenated matrix of size $MN\times P$, constructed by the column vector ${{\bf{\Xi }}_{i,p}}{\bf{x}}$ , i.e.,
\begin{equation}
{\bf{\Phi }}_i^{\omega _i^\tau ,\omega _i^{_\nu },{\bf{W}}_i^{{\rm{com}}}}\left( {\bf{x}} \right) = \left[ {{{\bf{\Xi }}_{i,1}}{\bf{x}}\quad {{\bf{\Xi }}_{i,2}}{\bf{x}}\quad  \cdots \quad {{\bf{\Xi }}_{i,P}}{\bf{x}}} \right],
\label{equation_phi}
\end{equation}
and ${{\bf{\Xi }}_{i,p}}$ is given by
\begin{equation}
{{\bf{\Xi }}_{i,p}} \buildrel \Delta \over = \left( {{{\bf{F}}_N} \otimes {{\bf{I}}_M}} \right){{\bm{\Pi }}^{{l_{i,p}}}}{{\bm{\Delta}} ^{{k_{i,p}}+{\kappa _{i,p}}}}{{\bf{W}}_{{a_{i,p}}}}\left( {{\bf{F}}_N^{\rm{H}} \otimes {{\bf{I}}_M}} \right).
\label{equation_Xi}
\end{equation}
To analyze the PEP performance for communication, we start from the study of the conditional  pairwise-error probability (PEP) based on~\eqref{new_com_model2}. In particular, let us define  the \emph{conditional Euclidean distance} ${d_{{\bf{h}}_i^{{\rm{eff}}},{{\bm{\omega}} _\tau },{{\bm{\omega}} _\nu },{{\bm \alpha} _i^{{\rm{com}}}},{\bf{W}}_i^{{\rm{com}}}}^2\left( {{\bf{x}},{\bf{x'}}} \right)}$ between a pair of codewords ${\bf{x}}$ and ${\bf{x'}}$ (${\bf{x}} \ne {\bf{x'}}$) by
\begin{align}
&d_{{\bf{h}}_i^{{\rm{eff}}},{\bm \omega} _i^\tau ,{\bm \omega}_i^{_\nu },{\bm \alpha} _i^{{\rm{com}}},{\bf{W}}_i^{{\rm{com}}}}^2\left( {{\bf{x}},{\bf{x'}}} \right)= d_{{\bf{h}}_i^{{\rm{eff}}},{\bm \omega} _i^\tau ,{\bm \omega}_i^{_\nu },{\bm \alpha} _i^{{\rm{com}}},{\bf{W}}_i^{{\rm{com}}}}^2\left( {\bf{e}} \right)
\buildrel \Delta \over ={\left\| {{\bf{\Phi }}_i^{{\bm \omega} _i^\tau ,{\bm \omega} _i^{_\nu },{\bf{W}}_i^{{\rm{com}}}}\left( {\bf{e}} \right){\bm \alpha} _i^{{\rm{com}}}{{\bf{h}}_i^{{\rm{eff}}}}} \right\|^2}\notag\\
=&{\left( {{\bf{h}}_i^{{\rm{eff}}}} \right)^{\rm{H}}} {\left({{\bm \alpha} _i^{{\rm{com}}}}\right)^{\rm H}}{\bf{\Omega }}_i^{_{{\bm \omega} _i^\tau ,{\bm \omega} _i^{_\nu },{\bf{W}}_i^{{\rm{com}}}}}\left( {\bf{e}} \right){{\bm \alpha} _i^{{\rm{com}}}}{\bf{h}}_i^{{\rm{eff}}}
={\left( {{\bf{h}}_i^{{\rm{eff}}}} \right)^{\rm{H}}}{\bf{\tilde \Omega }}_i^{{\bm \omega }_i^\tau ,{\bm \omega }_i^\nu,{\bf{W}}_i^{{\rm{com}}} }\left( {\bf{e}} \right){\bf{h}}_i^{{\rm{eff}}}, \label{too_long}
\end{align}
where ${\bf{e}} = {\bf{x}} - {\bf{x'}}$ is the corresponding codeword difference (error) sequence.
In particular, we refer to ${\bf{\Omega }}_i^{_{{\bm \omega} _i^\tau ,{\bm \omega} _i^{_\nu },{\bf{W}}_i^{{\rm{com}}}}}\left( {\bf{e}} \right) \buildrel \Delta \over = {\left( {{\bf{\Phi }}_i^{{\bm \omega} _i^\tau ,{\bm \omega} _i^{_\nu },{\bf{W}}_i^{{\rm{com}}}}\left( {\bf{e}} \right)} \right)^{\rm{H}}}{\bf{\Phi }}_i^{{\bm \omega} _i^\tau ,{\bm \omega} _i^{_\nu },{\bf{W}}_i^{{\rm{com}}}}\left( {\bf{e}} \right)
$ in~\eqref{too_long} as the \emph{codeword difference matrix}, while to ${\bf{\tilde \Omega }}_i^{_{{\bm \omega} _i^\tau ,{\bm \omega} _i^{_\nu },{\bf{W}}_i^{{\rm{com}}}}}\left( {\bf{e}} \right) \buildrel \Delta \over = {\left({{\bm \alpha} _i^{{\rm{com}}}}\right)^{\rm H}}{\bf{\Omega }}_i^{_{{\bm \omega} _i^\tau ,{\bm \omega} _i^{_\nu },{\bf{W}}_i^{{\rm{com}}}}}\left( {\bf{e}} \right){{\bm \alpha} _i^{{\rm{com}}}}$ as the \emph{weighted codeword difference matrix}.
For notational simplicity, we henceforth drop the superscript of ${\bf{\Omega }}_i^{_{{\bm \omega} _i^\tau ,{\bm \omega} _i^{_\nu },{\bf{W}}_i^{{\rm{com}}}}}\left( {\bf{e}} \right)$ and ${\bf{\tilde \Omega }}_i^{_{{\bm \omega} _i^\tau ,{\bm \omega} _i^{_\nu },{\bf{W}}_i^{{\rm{com}}}}}\left( {\bf{e}} \right)$, and we have
\begin{equation}
{\bf{\Omega }}_i\left( {\bf{e}} \right) = \left[ {\begin{array}{*{20}{c}}
{{{\bf{e}}^{\rm{H}}}{\bf{\Xi }}_{i,1}^{\rm{H}}{{\bf{\Xi }}_{i,1}}{\bf{e}}}&{{{\bf{e}}^{\rm{H}}}{\bf{\Xi }}_{i,1}^{\rm{H}}{{\bf{\Xi }}_{i,2}}{\bf{e}}}& \cdots &{{{\bf{e}}^{\rm{H}}}{\bf{\Xi }}_{i,1}^{\rm{H}}{{\bf{\Xi }}_{i,P}}{\bf{e}}}\\
{{{\bf{e}}^{\rm{H}}}{\bf{\Xi }}_{i,2}^{\rm{H}}{{\bf{\Xi }}_{i,1}}{\bf{e}}}&{{{\bf{e}}^{\rm{H}}}{\bf{\Xi }}_{i,2}^{\rm{H}}{{\bf{\Xi }}_{i,2}}{\bf{e}}}&{}& \vdots \\
 \vdots &{}& \ddots & \vdots \\
{{{\bf{e}}^{\rm{H}}}{\bf{\Xi }}_{i,P}^{\rm{H}}{{\bf{\Xi }}_{i,P}}{\bf{e}}}& \cdots & \cdots &{{{\bf{e}}^{\rm{H}}}{\bf{\Xi }}_{i,P}^{\rm{H}}{{\bf{\Xi }}_{i,P}}{\bf{e}}}
\end{array}} \right]. \label{Gram}
\end{equation}
According to~\cite{li2021performance}, the conditional PEP is upper-bounded by
\begin{align}
\Pr\left( {\left. {{\bf{x}},{\bf{x'}}} \right|{{\bf{h}}_i^{{\rm{eff}}}},{{\bm{\omega }}_i^\tau},{{\bm{\omega }}_i^\nu}},{\bf{W}}_i^{{\rm{com}}},{\bm \alpha} _i^{{\rm{com}}} \right) \le& \exp \left( { - \frac{{{1}}}{{4{N_0}}}{d_{{\bf{h}}_i^{{\rm{eff}}},{\bm \omega} _i^\tau ,{\bm \omega}_i^{_\nu },{\bm \alpha} _i^{{\rm{com}}},{\bf{W}}_i^{{\rm{com}}}}^2\left( {{\bf{x}},{\bf{x'}}} \right)}} \right)\notag\\
=& \exp \left( { - \frac{{{1}}}{{4{N_0}}}} {\left( {{\bf{h}}_i^{{\rm{eff}}}} \right)^{\rm{H}}}{\bf{\tilde \Omega }}_i\left( {\bf{e}} \right){\bf{h}}_i^{{\rm{eff}}}\right) .\label{PEP_derivation1}
\end{align}
To further simplify~\eqref{PEP_derivation1}, let us focus on the structures of both ${\bf{\tilde \Omega }}_i\left( {\bf{e}} \right)$ and ${\bf{\Omega }}_i\left( {\bf{e}} \right)$.
We observe that both ${\bf{\tilde \Omega }}_i\left( {\bf{e}} \right)$ and ${\bf{\Omega }}_i\left( {\bf{e}} \right)$ are positive-semidifinite Hermitian matrices by their definitions. Furthermore, with a proper design of power allocation, i.e., ${{\bm \alpha} _i^{{\rm{com}}}}$ is of full-rank, both ${\bf{\tilde \Omega }}_i\left( {\bf{e}} \right)$ and ${\bf{\Omega }}_i\left( {\bf{e}} \right)$ share the same rank.
Based on this observation, we consider the eigenvalue decomposition to further our derivation.
Let $r_i$ denote the rank of both ${\bf{\tilde \Omega }}_i\left( {\bf{e}} \right)$ and ${\bf{\Omega }}_i\left( {\bf{e}} \right)$, where $r_i \le P$.
Furthermore, let us denote by $\left\{ {{{\bf{u}}_{i,1}},{{\bf{u}}_{i,2}},...,{{\bf{u}}_{i,P}}} \right\}$ the eigenvectors of ${\bf{\tilde \Omega }}_i\left( {\bf{e}} \right)$ and
$\left\{ {{\lambda _i}\left[ 1 \right],{\lambda _i}\left[ 2 \right],...,{\lambda _i}\left[ P \right]} \right\}$ the corresponding nonnegative real eigenvalues sorted in the descending order,
where ${{\lambda _i\left[ j \right]}} > 0$ for $1 \le j \le r_i$ and ${{\lambda _i\left[ j \right]}} = 0$ for $r_i+1 \le j \le P$.
Then,~\eqref{PEP_derivation1} can be further expanded by~\cite{Tarokh1998space}
\begin{align}
\Pr\left( {\left. {{\bf{x}},{\bf{x'}}} \right|{{\bf{h}}_i^{{\rm{eff}}}},{{\bm{\omega }}_i^\tau},{{\bm{\omega }}_i^\nu}},{\bf{W}}_i^{{\rm{com}}},{\bm \alpha} _i^{{\rm{com}}} \right)& \le  \exp \left( { - \frac{1}{{4{N_0}}}\sum\limits_{j = 1}^{{r_i}} {{\lambda _i}\left[ j \right]} {{\left| {\bar h_i^{{\rm{eff}}}\left[ j \right]} \right|}^2}} \right), \label{PEP_derivation2}
\end{align}
where $\bar h_i^{{\rm{eff}}}\left[ j \right] = {{\bf{u}}_{i,j}}{\bf{h}}_i^{{\rm{eff}}}$, for $1 \le j \le r_i$.
Note that the exact values of the elements in ${\bf{h}}_i^{\rm{eff}}$ are unknown to the BS. Therefore, we need to consider the distributions of those elements in order to further our derivation.
It can be shown that $\left\{ {\bar h_i^{{\rm{eff}}}\left[ 1 \right],\bar h_i^{{\rm{eff}}}\left[ 2 \right],...,\bar h_i^{{\rm{eff}}}\left[ r_i \right]} \right\}$ are independent complex Gaussian random variables with zero mean and variance $1/(2P)$ per real dimension.
Consequently, ${\small| \bar h_i^{{\rm{eff}}}\left[ j \right] \small|}$ follows the Rayleigh distribution~\cite{Tarokh1998space}, whose
PDF is given by ${f_{{\rm{PDF}}}}\left( x \right) = 2Px\exp \left( { - P{{x}^2}} \right)$.
With the uniform power delay and Doppler profile, we can get rid of the influence of effective fading coefficients in~\eqref{PEP_derivation2} by averaging ${\small| \bar h_i^{{\rm{eff}}}\left[ j \right] \small|}$ term by term, yielding
\begin{align}
\Pr\left( {\left. {{\bf{x}},{\bf{x'}}} \right|{{\bm{\omega }}_i^\tau},{{\bm{\omega }}_i^\nu}},{\bf{W}}_i^{{\rm{com}}},{\bm \alpha} _i^{{\rm{com}}} \right) \le \prod\limits_{j = 1}^{{r_i}} {\frac{1}{{1 + \frac{{{\lambda _i[j]}}}{{4{N_0}P}}}}}  \le \frac{1}{{\prod\limits_{j = 1}^{{r_i}} {{\lambda _i[j]}} }}{\left( {\frac{1}{{4{N_0}P}}} \right)^{ - {r_i}}}. \label{PEP_derivation3}
\end{align}
As indicated by~\eqref{PEP_derivation3}, the PEP decreases exponentially with an order of ${r_i}$ with the reduction of the noise PSD. In fact, this exponent is the \textbf{diversity gain} of the transmission~\cite{Tarokh1998space,vucetic2003space,Yuan2003performance}.

In order to enable reliable transmissions, we aim to minimize the upper-bound in~\eqref{PEP_derivation3} by designing suitable precoding matrices and power allocation. To facilitate our design, let we first assume that there is a set of precoding matrices ${\bf{ W}}_i^{{\rm{com}}}$ and a specific power allocation ${\bm { \alpha} _i^{{\rm{com}}}}$ that can minimize the PEP upper-bound in~\eqref{PEP_derivation3} for any given $\bf e$ with respect to all possible delay and Doppler shifts, i.e., ${\bm{\omega }}_i^\tau$ and ${\bm{\omega }}_i^\nu$.
Then, we will develop practical precoding designs and power allocation in the next subsection, such that this lowest PEP upper-bound is approachable.
Notice that both ${\bf{\Omega }}_i\left( {\bf{e}} \right)$ and ${\bf{\tilde \Omega }}_i\left( {\bf{e}} \right)$ are \emph{Gram} matrices~\cite{Tut_Gram} of size $P \times P$ and thus the maximum value of the ranks of both ${\bf{\Omega }}_i\left( {\bf{e}} \right)$ and ${\bf{\Omega }}_i\left( {\bf{e}} \right)$ is $P$. In particular, when ${\bf{\Omega }}_i\left( {\bf{e}} \right)$ is of full-rank,~\eqref{PEP_derivation3} can be further simplified by
\begin{align}
\Pr\left( {\left. {{\bf{x}},{\bf{x'}}} \right|{{\bm{\omega }}_i^\tau},{{\bm{\omega }}_i^\nu}},{\bf{W}}_i^{{\rm{com}}},{\bm \alpha} _i^{{\rm{com}}} \right)& \le \frac{1}{{\prod\limits_{j = 1}^{{P}} {{\lambda _i[j]}} }}{\left( {\frac{1}{{4{N_0}P}}} \right)^{ - {P}}}= \frac{1}{{\det \left( {{{{\bf{\tilde \Omega }}}_i\left( {\bf{e}} \right)}} \right)}}{\left( {\frac{1}{{4{N_0}P}}} \right)^{ - P}}\notag\\
& = \frac{1}{{\det \left( {{{\bf{\Omega }}_i\left( {\bf{e}} \right)}} \right)}}{\left( {\frac{{{{\left( {\prod\limits_{p = 1}^P {{\alpha _{{a_{i,p}}}}} } \right)}^{\frac{1}{P}}}}}{{4{N_0}P}}} \right)^{ - P}}. \label{PEP_derivation4}
\end{align}
Based on~\eqref{PEP_derivation4}, we notice that the determinant of ${\bf{\Omega }}_i\left( {\bf{e}} \right)$ is related to the delay shifts ${{\bm{\omega }}_i^\tau}$, the Doppler shifts ${{\bm{\omega }}_i^\nu}$, and the precoding matrices ${\bf{W}}_i^{{\rm{com}}}$, but it is independent from the power allocation ${\bm \alpha} _i^{{\rm{com}}}$. This fact indicates that the we can analyze the influence from the precoding matrices and power allocation on the error performance separately.
In particular,~\eqref{PEP_derivation4} indicates that in the case of full-rank, the precoding scheme should be designed to maximize the determinant of the codeword difference matrix ${\bf{\Omega }}_i\left( {\bf{e}} \right)$.
To maximize the determinant of ${\bf{\Omega }}_i\left( {\bf{e}} \right)$, let us consider the following theorem.

\textbf{Theorem 1} (\emph{Upper-bound on the determinant of ${\bf{\Omega }}_i\left( {\bf{e}} \right)$}):
The determinant of the codeword difference matrix ${\bf{\Omega }}_i\left( {\bf{e}} \right)$ can be upper-bounded by
\begin{equation}
\det \left( {{{\bf{\Omega }}_i}\left( {\bf{e}} \right)} \right) \le {\left( {d_{\rm{E}}^2\left( {\bf{e}} \right)} \right)^P}, \label{determinant_upper_bound1}
\end{equation}
where $ {d_{\rm{E}}^2\left( {\bf{e}} \right)}$ denotes the Euclidean distance of the error sequence $\bf e$, i.e.,
$ {d_{\rm{E}}^2\left( {\bf{e}} \right)}  \buildrel \Delta \over = {\bf e}^{\rm H}{\bf e}$. Furthermore, a sufficient condition for achieving the equality is ${\bf{\Omega }}_i\left( {\bf{e}} \right)$ being a diagonal matrix.

\emph{Proof}: The proof is given in Appendix A.

In fact, Theorem~1 indicates that the PEP upper-bounded can be minimized if the received signals from different paths are orthogonal to each other~\cite{Li2020on,li2021performance}.
Notice that Theorem 1 provides an upper-bound of the determinant that is independent from the delay and Doppler shifts. By substituting~\eqref{determinant_upper_bound1} in~\eqref{PEP_derivation4}, we arrive at
\begin{align}
\Pr\left( \left. {{\bf{x}},{\bf{x'}}} \right|{\bf{W}}_i^{{\rm{com}}},{\bm \alpha} _i^{{\rm{com}}} \right)& \le  \frac{1}{{\left( {d_{\rm{E}}^2\left( {\bf{e}} \right)} \right)^P}}{\left( {\frac{{{{\left( {\prod\limits_{p = 1}^P {{\alpha _{{a_{i,p}}}}} } \right)}^{\frac{1}{P}}}}}{{4{N_0}P}}} \right)^{ - P}}={\left( {\frac{{d_{\rm{E}}^2\left( {\bf{e}} \right)}}{P}} \right)^{ - P}}{\left( {\frac{{{{\left( {\prod\limits_{p = 1}^P {{\alpha _{{a_{i,p}}}}} } \right)}^{\frac{1}{P}}}} }{{4{N_0}}}} \right)^{ - P}}.\label{PEP_derivation5}
\end{align}
According to~\eqref{PEP_derivation5}, we refer to the term ${d_{\rm{E}}^2\left( {\bf{e}} \right)}/P$ as the \textbf{maximum coding gain} of the underlying transmission~\cite{Li2020on,li2021performance}, which indicates how much can the error performance be possibly improved by varying $\bf e$, for all possible values of the delay and Doppler shifts.
In the next subsection, we will design suitable precoding matrices based on~\eqref{PEP_derivation5}.

\subsection{Precoding Design}
We notice that the delay and Doppler indices among different paths will affect the PEP performance.
Motivated by this observation, we propose our precoding design by considering the concepts of \textbf{virtual delay index} and \textbf{virtual Doppler index}, whose definitions are given as follows.

\textbf{Definition 1} (\emph{Virtual Delay and Doppler Indices}):
The virtual delay and Doppler indices are defined by $0 \le {{{\dot l}_p}} \le M-1$ and $0 \le {{{\dot k}_p}} \le N-1$, for $1 \le p \le P$, where ${{{\dot l}_p}}$ and ${{{\dot k}_p}}$ are of integer values, for $1 \le p, p' \le P$.

Recalling the discussions in previous subsections, we note that our precoding design is to shape the codeword difference matrix ${\bf{\Omega }}_i\left( {\bf{e}} \right)$, such that it can be a diagonal matrix, for any possible ${\bm{\omega }}_i^\tau$, ${\bm{\omega }}_i^\nu$, and $\bf e$.
By observing the structure of~\eqref{Gram} and according to Theorem 1, we notice that the aforementioned design criterion for precoding matrices is satisfied if
\begin{align}
{\bf e}^{\rm H}{\bf{\Xi }}_{_{i,p}}^{\rm{H}}{{\bf{\Xi }}_{i,p'}} {\bf e}= 0, \label{precoding_CR1}
\end{align}
and
\begin{align}
{\bf{W}}_{{a_{i,p}}}^{\rm{H}}{{\bf{W}}_{{a_{i,p}}}} = {{\bf{I}}_{MN}}, \label{precoding_CR2}
\end{align}
for any possible $\bf e$, and any $ 1 \le p,p' \le P$ and $p' \ne p$. Corresponding to both~\eqref{precoding_CR1} and~\eqref{precoding_CR2}, the following lemma shows an interesting fact of the precoding design problem.

\textbf{Lemma 3} (\emph{Determinant Dilemma}):
The precoding matrices cannot satisfy ${\bf{W}}_{{a_{i,j}}}^{\rm{H}}{{\bf{W}}_{{a_{i,p}}}} = {{\bf{I}}_{MN}}$ and ${\bf{W}}_{{a_{i,p}}}^{\rm{H}}{{\bf{W}}_{{a_{i,p'}}}} = {{\bf{0}}_{MN}}$ at the same time, for $ 1 \le p,p' \le P$ and $p' \ne p$.

\emph{Proof}: The proof is given in Appendix B.

As indicated by Lemma~3, we note that an explicit design algorithm of the precoding matrices satisfying the above criteria is not realizable.
Therefore, we consider a relaxation for the precoding design, where ${\bf{\Omega }}_i\left( {\bf{e}} \right)$ is a diagonally-dominant matrix~\cite{horn2012matrix} instead of a strict diagonal matrix.
To achieve this, we require that both the virtual delay and Doppler indices are different for different paths, i.e.,
${{{\dot l}_p}} \ne {{{\dot l}_{p'}}}$ and ${{{\dot k}_p}} \ne {{{\dot k}_{p'}}}$ for any $ 1 \le p,p' \le P$ and $p' \ne p$.
Therefore, for given \emph{a priori} AoA, delay and Doppler estimates from radar sensing, the proposed precoding matrices are of the form
\begin{align}
{{\bf{W}}_{{a_{i,p}}}} \buildrel \Delta \over = {{\bf{\Delta }}^{ - {{\hat k}_{i,p}} - {{\hat \kappa }_{i,p}}}}{{\bf{\Pi }}^{ - {{\hat l}_{i,p}}}}{{\bf{\Pi }}^{{{\dot l}_p}}}{{\bf{\Delta }}^{{{\dot k}_p}}}, \label{precoding_design}
\end{align}
where ${{\hat l}_{i,p}} $ and ${{{\hat k}_{i,p}} + {{\hat \kappa }_{i,p}}}$ are the estimated delay and Doppler indices, while ${{{\dot l}_p}}$ and ${{{\dot k}_p}}$ are the \emph{virtual} delay and Doppler indices with different values for different paths.

The rationale of the proposed precoding design is to improve the orthogonality devised by OTFS transmissions based on the nature of delay and Doppler shifts of the channel. Note that the delay and Doppler shifts are two physical parameters that determines the distortion characteristics of the resolvable path for transmission and different resolvable paths cannot share the same delay and Doppler shifts according to the definition~\cite{tse2005fundamentals,hlawatsch2011wireless}.
However, it is possible that different resolvable paths share the same delay or Doppler shift and in this case the natural orthogonality among different paths may be undermined. In the following proposition, we prove that the proposed precoding scheme can improve the orthogonality in the case where the delay or Doppler shifts associated to different paths are of the same values.

\textbf{Proposition 1} (\emph{Diagonal Dominance}):
In the case, where different paths share the same delay or Doppler indices, the codeword difference matrix ${\bf{\Omega }}_i\left( {\bf{e}} \right)$ is more likely to be a diagonally-dominant matrix with the proposed precoding.

\emph{Proof}: The proof is given in Appendix C.

\begin{figure}
\centering
\includegraphics[width=0.6\textwidth]{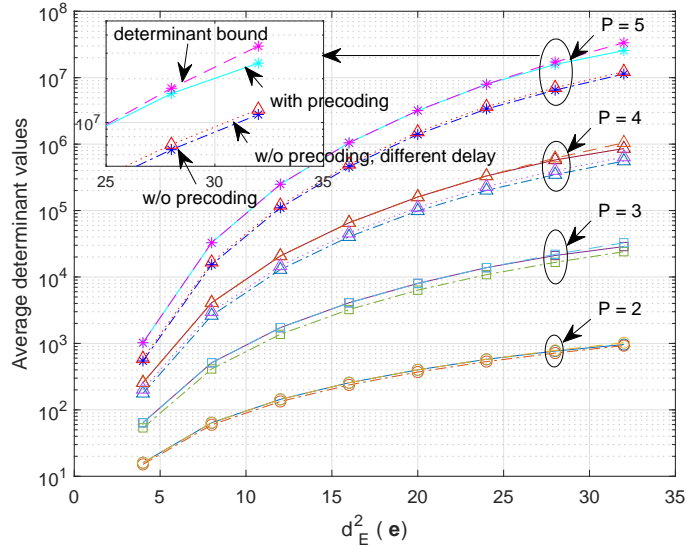}
\caption{Average determinant values of ${{\bf{\Omega }}_i\left( {\bf{e}} \right)}$ with (in dash-dotted lines) and without precoding (in solid lines for random delay indices and in dotted lines for different delay indices), comparing with the determinant bound in~\eqref{determinant_upper_bound1} (in dashed lines), where $M=8,N=8,l_{\rm max}=2$, and $k_{\rm max}=2$, respectively. }
\label{Determinant_evaluation}
\vspace{-5mm}
\centering
\end{figure}

Generally speaking, diagonally-dominant matrix are well-structured, whose determinant value is close to the corresponding diagonal matrix~\cite{horn2012matrix}.
To qualify the effectiveness of the proposed precoding, we compare the average determinant of ${{\bf{\Omega }}_i\left( {\bf{e}} \right)}$ with and without precoding by numerical simulations.
Without loss of generality, we consider $M=8,N=8$ in Fig.~\ref{Determinant_evaluation}, where the maximum delay and Doppler indices are set to be $l_{\rm{max}}=2$ and $k_{\rm{max}}=2$, respectively.
Since numerically emulating all the error sequences with such a frame size is generally intractable in a reasonable time frame even with BPSK mapping, we consider the comparison between the average determinant values of ${{\bf{\Omega }}_i\left( {\bf{e}} \right)}$ with and without precoding
for given error sequences\footnote{We use the same error sequence as in~\cite{li2021performance}, i.e., ${\bf e}=[2, 0, -2, 2, 0, -2,...,0...0]^{\rm T}$.}~\cite{Li2020on,li2021performance}, with respect to all possible channel realizations\footnote{Without loss of generality, we require the absolute value of the difference between any two Doppler indices no smaller than $0.2$.}.
For a better illustration, we also plot the determinant upper-bound in~\eqref{determinant_upper_bound1} for comparison.
As indicated by the figure, the proposed precoding can indeed increase the determinant value of ${{\bf{\Omega }}_i\left( {\bf{e}} \right)}$ compared to the case without precoding, where the determinant value with precoding aligns well with the upper-bound, especially for small values of $ {d_{\rm{E}}^2\left( {\bf{e}} \right)}$. In particular, we observe that the improvement becomes more obvious with more resolvable paths, which indicates that the proposed precoding is more helpful for communication transmissions in rich scattering scenarios.
On the other hand, it has reported in the literature that the fractional Doppler may potentially degrade the error performance~\cite{wei2020transmitter}. Therefore, we have also plotted the curves of the average determinant values corresponding to the case of different delay indices and fractional Doppler indices in Fig.~\ref{Determinant_evaluation}, for both $P=4$ and $P=5$. As indicated by the figure, the determinant values increase slightly compared to the non-precoded case (random delay and fractional Doppler indices), but still shows a noticeable gap compared to the precoded case. This observation agrees with the previous conclusions in~\cite{wei2020transmitter}.

\textbf{Remark 3}: It should also be noted that the effective DD domain channel matrix for the case of fractional Doppler shifts is generally dense and complex~\cite{Zhiqiang_magzine}. Consequently, the detection complexity required for fractional Doppler case is quite high~\cite{Raviteja2018interference}. Therefore, our proposed precoding design can also reduce the detection complexity at the UE side.

\textbf{Remark 4}: We briefly discuss the extension of our communication design in the case of inaccurate AoA estimates.
According to the proposed beam tracking scheme, it is suggested to allocate power to all the antennas in ${{\mathbb U}_{i,p}}$. Hence, given that the minor changes of the related parameters from the previous time instant are
well-compensated~\cite{liu2020joint}, we can apply the proposed precoding design to all the antennas in ${{\mathbb U}_{i,p}}$, and evenly assign the power among related antennas. We note that this will inevitably lead to
an SNR reduction for communication. However, this reduction is small if the beam width is small.
\subsection{Power Allocation for Communication}
As indicated by~\eqref{PEP_derivation5}, the power allocation should be designed to maximize the power product ${ {\prod\nolimits_{p = 1}^P {{\alpha _{{a_{i,p}}}}} } }$, i.e., the geometric mean of the allocated power associated to the corresponding paths in order to minimize the PEP. According to the
arithmetic mean-geometric mean (AM-GM) inequality, it is not hard to show that with a total power constraint, the optimal power allocation for minimizing PEP is the equal power allocation.
This is not unexpected, because the communication fading coefficients are of the same distribution and therefore the power allocation should not provide any bias to any path.

\textbf{Remark 5}: It is interesting to see that both radar sensing and communication require different power allocations.
Therefore, we in the following briefly discuss how to adapt those two allocations in practical systems.
Let us first consider the total power constraint for the $i$-th UE for the communication transmission, i.e., $\sum\nolimits_{p = 1}^P {{\alpha _{{a_{i,p}}}}} $. According to the AM-GM inequality, we have $\prod\nolimits_{p = 1}^P {{\alpha _{{a_{i,p}}}}}  \le {\left( {\frac{{\sum\nolimits_{p = 1}^P {{\alpha _{{a_{i,p}}}}} }}{P}} \right)^P}$. Notice that there is an exponent $P$ on the right hand side. Therefore, the communication performance degradation due to unfit power allocation may become more severe with a larger number of resolvable paths. However, it should be noted that the communication fading coefficients are not known at the BS. Therefore, the random nature of the communication channel may also mitigate the performance degradation induced by the undesirable power allocation.
On the other hand, since the proposed AoA estimation algorithm in Section III-B is based on the principle of matched-filtering, where the received power will be the key factor determining the estimation performance. Different from the communication counterpart, the radar reflection coefficients are assumed to be known at the BS. Therefore, a suitable power allocation that is designed specifically for each signal transmission can largely improve the sensing performance.
Therefore, in practical ISAC scenarios, radar sensing performance should be the priority for power allocation designs.

It should be noticed that a precise analysis on the relationship of the power allocation between radar sensing and communication performances requires detailed statistical models for both the radar reflection coefficients and communication fading coefficients, which is also related to system settings, e.g., the frame size. Since the major focus of this paper is to propose the ISAC transmission framework based on SS-OTFS modulation, we leave this interesting issue for our future work.

\section{Numerical Results}
We demonstrate the numerical results for the proposed ISAC transmissions in this section, where the sum-product algorithm (SPA) detection~\cite{li2020hybrid} for OTFS equalization is adopted at the UE side.
In specific, we set $N_{\rm BS}=128$, $M=32$, $N=16$, $\tau_{\max}=10$, and $\nu_{\max}=6$, respectively, unless specified otherwise, where the transmitted signals are BPSK  modulated.
To evaluate the communication performance, we define the average symbol SNR by $\frac{{{E_s}}}{{{N_0}}} \buildrel \Delta \over = \frac{\beta_i}{{{N_0}}}$, where $\beta_i$ denotes the average power assigned to each antenna for the $i$-th UE, i.e., ${\beta _i} = \sum\nolimits_{p = 1}^P {{\alpha _{{a_{i,p}}}}} /P$.
Meanwhile, we define the radar SNR by the ratio between the total power $\alpha_{\rm total}$ and the radar noise PSD ${\tilde N}_0$.
Without loss of generality, we assume that the reflection coefficient coefficients follow a uniform complex Gaussian distribution, i.e., ${{\tilde h}_{i,p}} \sim {\cal CN}\left( {0,1} \right)$.

\subsection{Beam Tracking Performance}
\begin{figure}
\centering
\includegraphics[width=0.5\textwidth]{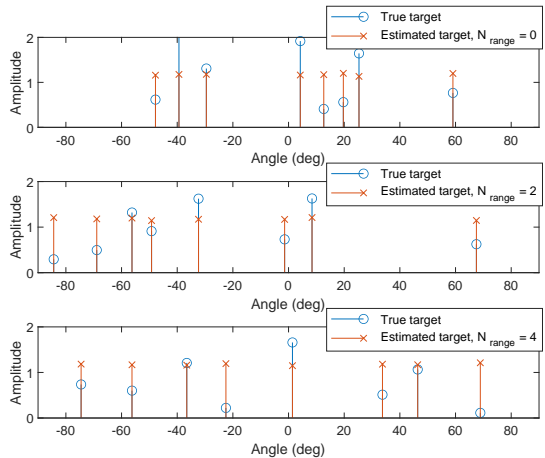}
\caption{AoA estimation performance with various beam widths for $K=4$ UEs, where each UE has $P=2$ paths.}
\label{AoA}
\vspace{-3mm}
\centering
\end{figure}

We present the AoA estimation performance for the proposed ISAC transmission with various beam widths in Fig.~\ref{AoA}, where we assume that there are in total $K=4$ UEs and each UE has $P=2$ paths. In specific, we consider a radar SNR at $5$ dB in Fig.~\ref{AoA}, where the proposed power allocation in Section III-C is applied and beam widths are controlled by the value of $N_{\rm range}$ as discussed in Section III-A. In particular, the amplitude in the figure represents the normalized trace of~\eqref{AoA_der2} with respect to the number of transmitted symbols.
As can be observed from the figure, the proposed beam tracking and AoA estimation can provide an accurate estimation performance for the considered scenario with various beam widths.
Note that the received echo power is reduced with the increase of the beam widths. In specific, it can be shown the average received power with $N_{\rm range}=4$ is only $1/5$ of that with $N_{\rm range}=0$, since we evenly assign the transmit power among all $N_{\rm range}+1$ antennas. Therefore, this observation also indicates that the proposed power allocation is suitable for the considered radar sensing issues.

\subsection{Precoding Performance}

\begin{figure}
\centering
\includegraphics[width=0.6\textwidth]{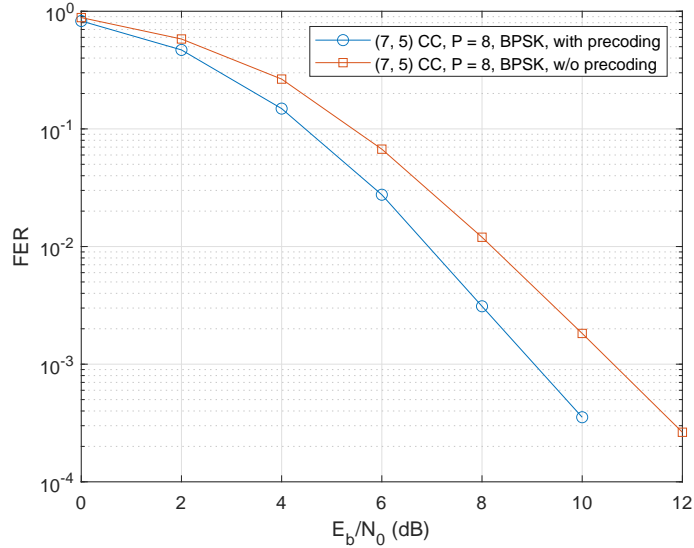}
\caption{FER comparisons between the proposed ISAC transmissions with and without precoding, where (7, 5) convolutionally coded BPSK signals are considered with $P = 8$.}
\label{Path8_coded}
\centering
\vspace{-3mm}
\end{figure}

We verify the effectiveness of the proposed precoding scheme in Fig.~\ref{Path8_coded}, where the FER performance for a specific UE with only integer delay and Doppler indices is illustrated. In particular, we consider the coded BPSK signals with $P=8$ for transmission. Due to the high detection/decoding complexity, we consider a smaller frame size for simulation, where we have $N=8$, $M=16$.
Without loss of generality, we apply the terminated (7, 5) convolutional code (CC) as the channel code, and the virtual delay and Doppler indices are randomly generated for precoding.
Meanwhile, we consider the equal power allocation in this example.
As can be observed from the figure, with the same channel coding and power allocation, the transmission with precoding has a roughly $1.7$ dB gain in terms of average bit SNR compared to the transmission without precoding at FER $4 \approx  \times 10^{-4}$.
Furthermore, we also notice that the FER slope for precoded transmission is steeper than the transmission without precoding. This observation indicates that the proposed precoding can also improve the diversity gain, which is due to the fact that the codeword difference matrix is more likely to have full-rank when different delay and Doppler indices are of different values.


\subsection{Radar Sensing and Communication Performances with Power Allocations}
As discussed in Remark 5, both radar sensing and communication requires different power allocations. Therefore, we evaluate the performances of radar sensing and communication with different power allocations in this subsection. In particular, we consider both the equal power allocation (designed for communication, referred to as ``w/o power allocation" in the figures) and the power allocation (designed for radar sensing, referred to as ``with power allocation" in the figures) proposed in Section III-C. Without loss of generality, we set $N_{\rm range}=0$.

\begin{figure}
\centering
\includegraphics[width=0.5\textwidth]{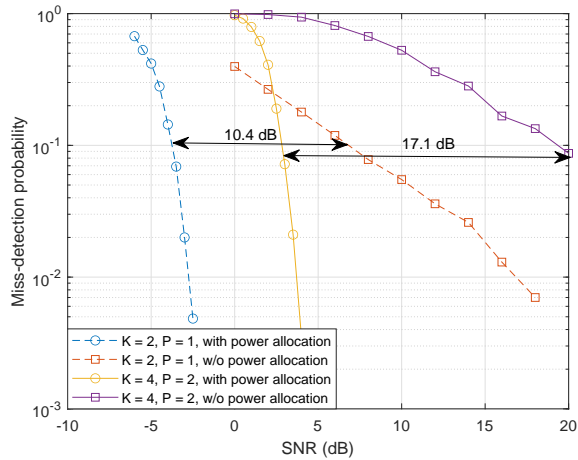}
\caption{Miss-detection probability for radar sensing with and without power allocation for different number of UEs and paths.}
\label{Misdetection}
\vspace{-3mm}
\centering
\end{figure}

To demonstrate the advantages of the proposed ISAC transmission, we consider a strict performance metric for radar sensing. As the angular domain is discretized, we are interested in the ``miss-detection probability" which is defined by the ratio between the times when the radar does not accurately detect the receive antenna indices ${\tilde a}_{i,p}$ for $1 \le i \le K$ and $1\le p\le P$ and the total number of ISAC transmissions. Without loss of generality, we consider two cases for radar sensing in Fig.~\ref{Misdetection}, where the number of UEs are $K=4$ and $K=2$, and the number of paths are $P=2$ and $P=1$, respectively. As indicated by the figure, suitable power allocation can provide significant performance improvements for radar sensing, especially when the number of targets is large. This indicates that the proposed power allocation is indeed suitable for radar sensing.

\begin{figure}
\centering
\includegraphics[width=0.5\textwidth]{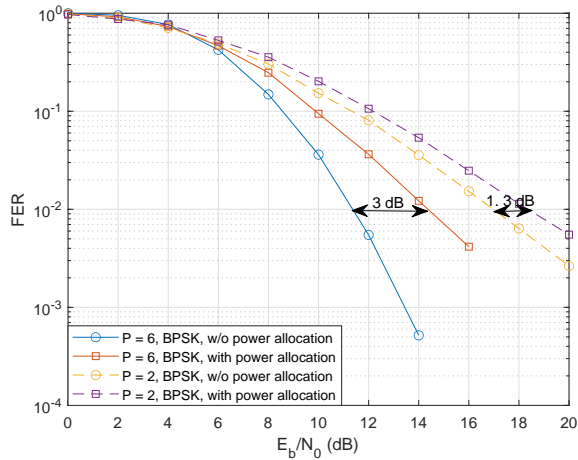}
\caption{Comparisons between the FER performances for a specific UE with equal power allocation and with the power allocation given in Section III-C.}
\label{FER_PA}
\vspace{-3mm}
\centering
\end{figure}
We show the FER performances for a specific UE with different power allocations for communications in Fig.~\ref{FER_PA}, where the proposed precoding scheme is applied. As shown in the figure, equal power allocation can provide a better error performance, and the performance improvement becomes larger with more paths, which are consistent with our discussions in Section IV-C. On the other hand, we also notice that with the power allocation designed for radar sensing, the communication performance become worse. However, this performance degradation is relatively small compared to the performance improvement for radar sensing. Therefore, it is desirable to design the power allocation with a priority for improving the radar sensing performance.

\section{Conclusion}
In this paper, we proposed a novel framework for ISAC transmissions based on the SS-OTFS modulation by considering the mismatch of the reflection strengths between the radar sensing and communication. We first derived the
channel models for both radar sensing and communication, which are then simplified based on the properties of SS-OTFS modulation.
Based on the radar model, we proposed simple beam tracking, AoA estimation algorithms, and power allocation for radar sensing. Furthermore, we carried out a detailed analysis on the
PEP for communication, where we showed that the PEP can be minimized with equal power allocation if the received signals from different paths are orthogonal to each other.
Based on this conclusion, we proposed a symbol-wise precoding design to improve this orthogonality by introducing virtual delay and Doppler indices. We also noticed that radar sensing and communication require different power allocations.
To facilitate the power allocation design for practical ISAC systems, we briefly discussed the radar sensing and communication performances with different power allocations and concluded
that the power allocation should be designed leaning towards radar sensing.
Finally, the effectiveness of the proposed framework is verified by numerical results.

\appendices

\section{Proof of Theorem 1}
According to~\eqref{Gram}, we notice that the codeword difference matrix ${{\bf{\Omega }}_i\left( {\bf{e}} \right)}$ is a Gram matrix corresponding to the vectors $\left\{ {{{\bf{{u}}}_{i,1}},{{\bf{{u}}}_{i,2}}, \ldots {{\bf{{u}}}_{i,P}}} \right\}$, where ${{\bf{u}}_{i,j}} \buildrel \Delta \over = {{\bf{\Xi }}_{i,j}}{\bf{e}}$.
In particular, the determinant of the Gram matrix ${{\bf{\Omega }}\left( {\bf{e}} \right)}$, i.e., the Gram determinant, is equal to the square of the $P$-dimensional volume of the parallelotope constructed on $\left\{ {{{\bf{{u}}}_{i,1}},{{\bf{{u}}}_{i,2}}, \ldots {{\bf{{u}}}_{i,P}}} \right\}$.
Let us refer to ${\rm{GD}}\left(\left\{ {{{\bf{{u}}}_{i,1}},{{\bf{{u}}}_{i,2}}, \ldots {{\bf{{u}}}_{i,P}}} \right\} \right)$ as the Gram determinant of ${{\bf{\Omega }}_i\left( {\bf{e}} \right)}$. It can be shown that the Gram determinant can be calculated recursively, such as \cite{Tut_Gram}
\begin{equation}
{\rm{GD}}\left( \left\{ {{{\bf{{u}}}_{i,1}},{{\bf{{u}}}_{i,2}}, \ldots {{\bf{{u}}}_{i,j}}} \right\} \right) = {\rm{GD}}\left( \left\{ {{{\bf{{u}}}_{i,1}},{{\bf{{u}}}_{i,2}}, \ldots {{\bf{{u}}}_{i,j-1}}} \right\} \right){\left\| {{{{\bf{\tilde u}}}_{i,j}}} \right\|^2},     \label{Gram_derivation1}
\end{equation}
where the term ${{{{\bf{\tilde u}}}_{i,j}}}$ denotes the \emph{orthogonal projection} of ${{{{\bf{u}}}_{i,j}}}$ onto the \emph{orthogonal complement} of ${\rm{span}}\left( {{{{\bf{ u}}}_{i,1}},{{{\bf{ u}}}_{i,2}}, \ldots {{{\bf{ u}}}_{i,j - 1}}} \right)$.
Considering the property of orthogonal projection, we have
\begin{align}
{\rm{GD}}\left(\left\{ {{{\bf{{u}}}_{i,1}},{{\bf{{u}}}_{i,2}}, \ldots {{\bf{{u}}}_{i,j}}} \right\} \right)& = {\rm{GD}}\left( \left\{ {{{\bf{{u}}}_{i,1}},{{\bf{{u}}}_{i,2}}, \ldots {{\bf{{u}}}_{i,j-1}}} \right\} \right){\left\| {{{{\bf{\tilde u}}}_{i,j}}} \right\|^2} \notag\\
&\le {\rm{GD}}\left( {\left\{ {{{\bf{{u}}}_{i,1}},{{\bf{{u}}}_{i,2}}, \ldots {{\bf{{u}}}_{i,j-1}}} \right\}} \right){\left\| {{{{\bf{u}}}_{i,j}}} \right\|^2},\label{Gram_derivation2}
\end{align}
where the equality holds if ${{{\bf{u}}_{i,j}}}$ is orthogonal to ${{{\bf{u}}_{i,j'}}}$, for $1\le j' < j$, i.e., ${\left( {{{\bf{\Xi }}_{i,j'}}{\bf{e}}} \right)^{\rm{H}}}\left( {{{\bf{\Xi }}_{i,j}}{\bf{e}}} \right) = 0$, for $1\le j' < j$.
Hence, by considering (\ref{Gram_derivation2}), the Gram determinant can be upper-bounded by
\begin{align}
\det \left( {{\bf{\Omega }}_i\left( {\bf{e}} \right)} \right) &= \prod\limits_{j = 1}^{P - 1} {{\rm{GD}}\left( {\left\{ {{{\bf{{u}}}_{i,1}},{{\bf{{u}}}_{i,2}}, \ldots {{\bf{{u}}}_{i,j}}} \right\}} \right){{\left\| {{{{\bf{\tilde u}}}_{i,j+1}}} \right\|}^2}}
 \le \prod\limits_{j = 1}^{P - 1} {{\rm{GD}}\left( {\left\{ {{{\bf{{u}}}_{i,1}},{{\bf{{u}}}_{i,2}}, \ldots {{\bf{{u}}}_{i,j}}} \right\}} \right){{\left\| {{{{\bf{u}}}_{i,j+1}}} \right\|}^2}} \notag\\
&\le \prod\limits_{j = 1}^P {{{\left\| {{{{\bf{ u}}}_{i,j}}} \right\|}^2}}
=\prod\limits_{j = 1}^{P - 1} {{\bf{e}}^{\rm{H}}}\left( {{{\bf{F}}_N} \otimes {{\bf{I}}_M}} \right){\bf{W}}_{{a_{i,j}}}^{\rm{H}}{{\bf{W}}_{{a_{i,j}}}}\left( {{\bf{F}}_N^{\rm{H}} \otimes {{\bf{I}}_M}} \right){\bf{e}}.\label{Gram_derivation3}
\end{align}
By noticing that ${\bf{W}}_{{a_{i,j}}}^{\rm{H}}{{\bf{W}}_{{a_{i,j}}}}= {\bf I}_{MN}$,~\eqref{Gram_derivation3} can be further derived as~\eqref{determinant_upper_bound1}.
This completes the proof of Theorem 1.
\section{Proof of Lemma 3}
Assuming that ${\bf{W}}_{{a_{i,p}}}^{\rm{H}}{{\bf{W}}_{{a_{i,p'}}}} = {{\bf{0}}_{MN}}$, for $ 1 \le p,p' \le P$ and $p' \ne p$. Then, it is obvious that there exists an index $p$, where $1 \le p \le P$, such that ${\bf{W}}_{{a_{i,p}}}$ has a zero determinant, which is contradict to ${\bf{W}}_{{a_{i,p}}}^{\rm{H}}{{\bf{W}}_{{a_{i,p}}}} = {{\bf{I}}_{MN}}$.
This completes the proof of Lemma~3.
\section{Proof of Proposition 1}
To prove Proposition 1, we are interested in the absolute values of the non-diagonal elements codeword difference matrix ${{{\bf{\Omega }}_i}\left( {\bf{e}} \right)}$.
In particular, we will show that the absolute values of the non-diagonal elements are more likely to be reduced in the case where different paths share the same delay or Doppler shifts, even with the same error sequence $\bf e$.
To this end, let us focus on the $(p,p')$-th element of ${{\bf{\Omega }}_i}\left( {\bf{e}} \right)$, and it is rewritten by
\begin{align}
{{\bf{e}}^{\rm{H}}}{\bf{\Xi }}_{i,p}^{\rm{H}}{{\bf{\Xi }}_{i,p'}}{\bf{e}} =& {{\bf{e}}^{\rm{H}}}\left( {{{\bf{F}}_N} \otimes {{\bf{I}}_M}} \right){{\bm \Delta} ^{ - {{ k}_p}-{\kappa _p}}}{{\bm \Pi} ^{ - {{ l}_p}}}{{\bm \Pi} ^{{{ l}_{p'}}}}{{\bm \Delta} ^{{{ k}_{p'}+{\kappa _{p'}}}}}\left( {{\bf{F}}_N^{\rm{H}} \otimes {{\bf{I}}_M}} \right){\bf{e}}\notag\\
 = &{{{\bf{\tilde e}}}^{\rm{H}}}{{\bm\Delta} ^{ - {{ k}_p}-{\kappa _{p}}}}{{\bm\Pi} ^{{{ l}_{p'}} - {{ l}_p}}}{{\bm\Delta} ^{{{ k}_{p'}+{\kappa _{p'}}}}}{\bf{\tilde e}},
\label{Pro1_der1}
\end{align}
where ${\bf{\tilde e}} \buildrel \Delta \over = \left( {{\bf{F}}_N^{\rm{H}} \otimes {{\bf{I}}_M}} \right){\bf{e}}$ is the TDA domain error sequence.
Notice that each TDA domain OTFS symbol is a superposition of $N$ DD domain OTFS symbols with specific phase rotations according to the spreading effect of IFFT. Thus, in practical systems with a sufficiently large $N$, the TDA domain OTFS symbols behave like i.i.d. Gaussian variable due to the law of large numbers~\cite{li2021cross}. Let $n' = {\left[ {n - 1 - \left( {{l_{p'}} - {l_p}} \right)} \right]_{MN}} + 1$. Then,~\eqref{Pro1_der1} can be further simplified by
\begin{align}
{{\bf{e}}^{\rm{H}}}{\bf{\Xi }}_{i,p}^{\rm{H}}{{\bf{\Xi }}_{i,p'}}{\bf{e}} =\sum\limits_{n = 1}^{MN} {{e^{j\frac{{2\pi }}{{MN}}\left( {\left( {n' - 1} \right)\left( {{k_{p'}} + {\kappa _{p'}}} \right) - \left( {n - 1} \right)\left( {{k_p} + {\kappa _p}} \right)} \right)}}{{\tilde e}^*}\left[ n \right]\tilde e\left[ {n'} \right]}. \label{Pro1_der2}
\end{align}
In the case where $l_p=l_{p'}$, i.e., different paths share the same delay index,~\eqref{Pro1_der2} is simplified by
\begin{align}
{{\bf{e}}^{\rm{H}}}{\bf{\Xi }}_{i,p}^{\rm{H}}{{\bf{\Xi }}_{i,p'}}{\bf{e}}= {{{\bf{\tilde e}}}^{\rm{H}}}{{\bm\Delta} ^{ {{ k}_{p'}+{\kappa _{p'}}}- {{ k}_p-{\kappa _{p}}}}}{\bf{\tilde e}}=\sum\limits_{n = 1}^{MN} {{e^{j\frac{{2\pi }}{{MN}}\left( {n - 1} \right)\left( {{k_{p'}} + {\kappa _{p'}} - {k_p} - {\kappa _p}} \right)}}{{\left| {\tilde e\left[ n \right]} \right|}^2}} .
\label{Pro1_der3}
\end{align}
Comparing~\eqref{Pro1_der2} and~\eqref{Pro1_der3}, we observe that the absolute value of~\eqref{Pro1_der3} is more likely to be larger than that of~\eqref{Pro1_der2}, because the expectation of the term ${{{\tilde e}^*}\left[ n \right]\tilde e\left[ {n'} \right]}$ is zero based on the i.i.d. assumption~\cite{li2021cross}, while the term ${{\left| {\tilde e\left[ n \right]} \right|}^2}$ is strictly non-negative.
On the other hand, when ${{k_p} + {\kappa _p}}={{k_{p'}} + {\kappa _{p'}}}$, i.e., different paths share the same Doppler index,~\eqref{Pro1_der2} is simplified by
\begin{align}
{{\bf{e}}^{\rm{H}}}{\bf{\Xi }}_{i,p}^{\rm{H}}{{\bf{\Xi }}_{i,p'}}{\bf{e}}= \sum\limits_{n = 1}^{MN} {{e^{j\frac{{2\pi }}{{MN}}\left( {n' - n} \right)\left( {{k_p} + {\kappa _p}} \right)}}{{\tilde e}^*}\left[ n \right]\tilde e\left[ {n'} \right]}  \approx {e^{j\frac{{2\pi }}{{MN}}\left( {{l_p} - {l_{p'}}} \right)\left( {{k_p} + {\kappa _p}} \right)}}\sum\limits_{n = 1}^{MN} {{{\tilde e}^*}\left[ n \right]\tilde e\left[ {n'} \right]}  ,
\label{Pro1_der4}
\end{align}
whose absolute value is given by $\left| {\sum\limits_{n = 1}^{MN} {{{\tilde e}^*}\left[ n \right]\tilde e\left[ {n'} \right]} } \right|$. Therefore, it can be shown that the absolute value of~\eqref{Pro1_der2} is no larger than $\left| {\sum\limits_{n = 1}^{MN} {{{\tilde e}^*}\left[ n \right]\tilde e\left[ {n'} \right]} } \right|$, because $\left| {\sum\limits_{n = 1}^{MN} {a_n b_n} } \right| \le \sum\limits_{n = 1}^{MN} {\left| a_n \right|\left| b_n \right|} $, for any two arbitrary complex vectors $\bf a$ and $\bf b$.
This completes the proof of Proposition~1.

\bibliographystyle{IEEEtran}
\bibliography{ISAC}

\end{document}